\documentclass{elsart}
\usepackage{graphicx}
\usepackage{epsfig}
\usepackage{amssymb}

\begin{document}
 \begin{frontmatter}

 \title{A Simple Statistical Mechanical Approach for Studying Multilayer Adsorption of
Interacting Polyatomics}

\author[label1,label2]{G. D. Garc\'{\i}a},
\author[label1,label2]{F. O. S\'anchez-Varretti},
\author[label1,label3]{F. Rom\'a},
\author[label1]{A. J. Ramirez-Pastor\corauthref{cor1}}

\address[label1]{Dpto. de F\'{\i}sica, Instituto de F\'{\i}sica Aplicada, Universidad Nacional de San Luis - CONICET, Chacabuco 917, 5700 San Luis, Argentina.}
\address[label2]{Universidad Tecnol\'ogica Nacional, Regional San Rafael, Gral. Urquiza 314, 5600, San Rafael, Mendoza, Argentina. }
\address[label3]{Centro At\'omico Bariloche, San Carlos de Bariloche, R\'{\i}o Negro, R8402AGP, Argentina}
\thanks[cor1]{Corresponding author. Fax +54-2652-430224, E-mail: antorami@unsl.edu.ar}

\begin{abstract}
A simple statistical mechanical approach for studying multilayer
adsorption of interacting polyatomic adsorbates ($k$-mers) has
been presented. The new theoretical framework has been developed
on a generalization in the spirit of the lattice-gas model and the
classical Bragg-Williams (BWA) and quasi-chemical (QCA)
approximations. The derivation of the equilibrium equations allows
the extension of the well-known Brunauer-Emmet-Teller (BET)
isotherm to more complex systems. The formalism reproduces the
classical theory for monomers, leads to the exact statistical
thermodynamics of interacting $k$-mers adsorbed in one dimension,
and provides a close approximation for two-dimensional systems
accounting multisite occupancy and lateral interactions in the
first layer. Comparisons between analytical data and Monte Carlo
simulations were performed in order to test the validity of the
theoretical model. The study showed that: $(i)$ the resulting
thermodynamic description obtained from QCA is significantly
better than that obtained from BWA and still mathematically
handable; $(ii)$ for non-interacting $k$-mers, the BET equation
leads to an underestimate of the true monolayer volume; $(iii)$
attractive lateral interactions compensate the effect of the
multisite occupancy and the monolayer volume predicted by BET
equation agrees very well with the corresponding true value; and
$(iv)$ repulsive couplings between the admolecules hamper the
formation of the monolayer and the BET results are not good (even
worse than those obtained in the non-interacting case).
\end{abstract}
\end{frontmatter}

 \newpage
 \section{Introduction}

Adsorption on solid surfaces is a complex phenomenon
\cite{Hill,Clark,Steele,Adamson,Rudzi}, which implies a series of
questions about the nature of the forces binding foreign molecules
to a surface and about the thermodynamic behavior of the system.
Although the problem of the forces is still far from being
elucidated, a large amount of interest has been devoted to the
study of statistical properties, within the frame of simplified
models for both monolayer and multilayer adsorption. In 1918
Langmuir derived the monolayer adsorption isotherm kinetically for
gas molecules adsorbed on the homogeneous surface of adsorbents
without attractions among the adsorbed molecules \cite{Langmuir}.
After that Brunauer, Emmett and Teller \cite{BET}, based on a
model of localized adsorption, developed the most important theory
of multilayer adsorption. Their equation, the BET isotherm, was
the first and the most useful, covering the complete range of
pressures up to $p_0$, the saturation pressure. Later, Hill
\cite{Hill} derived the BET isotherm statistically on a group of
homogeneous adsorption sites for the multilayer adsorption since
it was derived kinetically by Brunauer, Emmett and Teller. It is
found to be in good agreement with some experimental data for
relative pressures less than about $0.5$ \cite{Gregg}.

But the theoretical BET isotherm deals with the common assumption
that each ad-molecule occupies one adsorption site of the surface.
However, most adsorbates involved in experiments are polyatomic;
hence, the theoretical description of their thermodynamic
properties is a topic of much interest in adsorption
theory\footnote{Even for simple gases such as oxygen, nitrogen and
carbon monoxide, which basically are not altered in their
molecular dimensions under physical adsorption, the adsorption
energy depends in general on the orientation of the molecule in
the adsorbed state.}. Leading contributions to this subject,
generically called multisite-occupancy adsorption, have been
presented by  Flory \cite{FLORY}, Huggins \cite{HUGGI}, Guggenheim
\cite{GUGGE}, DiMarzio \cite{DIMA}, Nitta et al. \cite{Nitta} and
Rudzinski et al. \cite{Rudzi1} through approximate treatments of
monolayer adsorption on homogeneous and heterogeneous surfaces.
More recently, Aranovich and Donohue \cite{ARA1,ARA2} derived a
multilayer adsorption isotherm, which is not limited by the
functional form of the monolayer adsorption isotherm and should be
capable to include multisite occupancy (with an adequate choice of
a fitting parameter). On the other hand, the closed exact solution
for the multilayer adsorption isotherm of dimers, along with the
basis for calculating adsorption thermodynamics of homonuclear
polyatomic molecules ($k$-mers) on one-(1D) and two-dimensional
(2D) substrates, have been recently presented \cite{LANG8,SS10}.
This rigorous thermodynamic study demonstrated that the entropic
contribution of non-spherical adsorbates is significant in the
multilayer regime when compared with monoatomic adsorption. Thus,
the determinations of surface areas and adsorption energies from
polyatomic adsorbate adsorption may be severely misestimated, if
this polyatomic character is not properly incorporated in the
thermodynamic functions from which experiments are interpreted.

There is another important physical fact which has not been
sufficiently studied; namely, the effect of the lateral
interactions between the ad-molecules in presence of multilayer
adsorption and multisite occupancy. Thus, most of the theoretical
work dealing with adsorption of interacting polyatomics has been
based on models of monolayer adsorption. In this context, the aim
of the present work is to extend the treatment of Refs.
\cite{LANG8,SS10} to include lateral interactions in the
adsorbate. Here, we introduce nearest-neighbor interactions
between the molecules adsorbed in the first layer\footnote{ As it
is well-known, BET equation can be applied at coverage not greatly
exceeding (statistically) monolayer coverage. Thus, although the
contribution from the secondary adsorption can already be
significant, the density of the molecules in the second and higher
adlayers is expected to be much lower than that in the first
adsorbed layer. Therefore, it seems to be satisfactorily enough to
take into account only the interactions between the primarily
adsorbed molecules \cite{Hill}.}, by following the
configuration-counting procedure of the Bragg-Williams approach
and the quasi-chemical approximation. In addition, Monte Carlo
(MC) simulations are performed in order to test the validity of
the theoretical model. The new theoretical scheme allows us: (1)
to reproduce the classical theory for monomers \cite{Hill,Rudzi};
(2) to develop a closed exact expression for the multilayer
adsorption isotherm of interacting $k$-mers on 1D chains; (3) to
obtain an accurate approximation for multilayer adsorption on 2D
substrates accounting multisite occupancy and lateral
interactions; and (4) to provide a simple model from which
experiments may be reinterpreted.

The present work is organized as follows. In Section 2, the
theoretical formalism along with the basis of the MC method are
presented. In Section 3, the results of the theoretical model are
shown and discussed by comparing with MC simulations. Finally,
conclusions are drawn in Section 4.

 \section{Basic Definitions: Adsorption Model and
Monte Carlo Simulation}

 \subsection{Model}

In this section we present the lattice-gas model for the
adsorption of particles with multisite occupancy in the multilayer
regime. The adsorbent is a homogeneous lattice of sites with
coordination number $\zeta$. The adsorbate is assumed as linear
molecules having $k$-identical units ($k$-mers) each of which
occupies an adsorption site. Furthermore, i) a $k$-mer can adsorb
exactly onto an already adsorbed one; ii) attractive and repulsive
lateral interactions are considered in the first layer and
horizontal interactions are ignored in higher layers; iii) the
adsorption heat in all layers, except the first one, equals the
molar heat of condensation of the adsorbate in bulk liquid phase.
Thus, $c=q_{1}/q_{i}=q_{1}/q$ with $q_{i}=q \ \ \
(i=2,...,\infty)$ denotes the ratio between  the single-molecule
partition functions in the first and higher layers \cite{LANG8}.
The fact that $k$-mers can arrange in the first layer leaving
sequences of $l$ empty sites with $l<k$, where no further
adsorption of a $k-$mer can occur in such a configuration, makes
the calculation of entropy  much elaborated than the one for
monomer adsorption.

To describe a system of $N$ $k$-mers adsorbed on $M$ sites at a
given temperature $T$, let us introduce the occupation variable
$s_i$ which can take the values $s_i=0$ or $1$, if the site $i$ is
empty or occupied by a $k$-mer unit, respectively. Under these
conditions, the Hamiltonian of the system is given by
\begin{equation}
H= \sum _{i=1} ^M \epsilon_{1} s_i  + k \epsilon \left(N-N_1
\right) + \sum _{\langle i,j \rangle} w s_i s_j - N_1(k-1)w
\label{h},
\end{equation}
where $\epsilon_{1}$ ($\epsilon$) represents the adsorption energy
of a $k$-mer unit on the first layer (higher layers); $N_1=\sum
_{i=1} ^M s_i/k$ is the number of $k$-mers adsorbed on the first
layer; $w$ is the lateral interaction energy between two
nearest-neighbor (NN) units belonging to different $k$-mers
adsorbed in the first layer (we use $w>0$ for repulsive and $w<0$
for attractive interactions) and $\langle i,j \rangle$ represents
pairs of NN sites. The term $N_1(k-1)w$ is subtracted in equation
(\ref{h}) since the summation over all the pairs of NN sites
overestimates the total energy by including $N_1(k-1)$ bonds
belonging to the $N_1$ adsorbed $k$-mers.

\subsection{Theory}

From a theoretical point of view, when intermolecular forces are
introduced (in our case, NN interactions in the first layer), an
extra term in the partition function for interaction energy is
required. With this extra term, only partition functions for the
whole system can be written. Ising \cite{Ising} gave an exact
solution to the 1D monolayer in 1925. All other cases are
expressed in terms of series solution \cite{Hill,Domb}, except for
the special case of 2D monolayers at half-coverage, which was
exactly solved by Onsager \cite{Onsager} in 1944. Close
approximate solutions in dimensions higher than one can be
obtained, and the two most important of these are the
Bragg-Williams approximation (BWA) \cite{Hill} and the
quasi-chemical approximation (QCA) \cite{Hill,Bethe}. These
leading models have played a central role in the study of
adsorption systems in presence of lateral interactions between the
adatoms. Next, we apply BWA and QCA to study multilayer adsorption
of interacting $k$-mers.

The BWA is the simplest mean-field treatment for interacting
adsorbed particles, even in the case of multilayer adsorption and
multisite occupancy. For a lattice having $M$ adsorption sites,
the maximum number of columns that can be grown up onto it is
$n_{max}=M/k$. If an infinite number of layers is allowed to
develop on the surface, the grand partition function, $\Xi$, of
the adlayer in equilibrium with a gas phase at chemical potential
$\mu$ and temperature $T$, is given by
\begin{equation}
\Xi=\sum_{n=0}^{n_{max}} \xi^{n} \Omega_{k}(n,M,\zeta)
\exp{\left[-\beta \overline{E_k}(n,M) \right]} \label{xi},
\end{equation}
where $\xi$ is the grand partition function of a single column of
$k$-mers having at least one $k$-mer in the first layer;
$\Omega_{k}(n,M,\zeta)$ is the total number of distinguishable
configurations of $n$ columns on $M$ sites with connectivity
$\zeta$ and $\overline{E_k}(n,M)$ is the mean total energy of the
system assuming that the $n$ columns are randomly distributed over
the lattice.

Then,
\begin{equation}
\xi = \sum_{i=1}^{\infty} q_{1} q^{i-1} \lambda^{i }= c
\sum_{i=1}^{\infty} q^{i} \lambda^{i} = {{c \lambda q} \over
{1-\lambda q}} = {{c x} \over {1-x}} \label{xi1},
\end{equation}
where $\lambda=\exp(\mu/k_{B}T)$ is the fugacity and $k_B$ is the
Boltzmann constant. In addition, it is possible to demonstrate
that $x=\lambda q =p/p_o$ is the relative pressure
\cite{Rudzi,LANG8}.

$\Omega_k(n,M,\zeta)$ can be approximated considering that the
columns are distributed completely at random on the lattice and
assuming the arguments given by different authors
\cite{Flory1,Flory2,PRB3} to relate the configurational factor
$\Omega_k(n,M,\zeta)$ for any $\zeta$, with the same quantity in
the 1D case $(\zeta=2)$. Thus
\begin{equation}
\Omega_k(n,M,\zeta)=\eta(\zeta,k)^n \Omega_k(n,M,2),
\end{equation}
where $\eta(\zeta,k)$ is, in general, a function of the
connectivity and the size of the molecules. In the particular case
of rigid straight $k$-mers it follows that
$\eta(\zeta,k)=\zeta/2$. In addition, $\Omega_k(n,M,2)$ can be
readily calculated \cite{PRB3} giving
\begin{equation}
\Omega_k(n,M,2) = {\left[M-\left(k-1 \right) n \right] ! \over n !
\left[M-kn \right] !}.
\end{equation}

On the other hand, $\overline{E_k}(n,M)$ needs to be calculated.
For this purpose, let us consider the number of NN of a $k$-mer
(column) adsorbed in the first layer,
\begin{equation}
z=\left[ 2(\zeta-1)+(k-2)(\zeta-2)\right] \label{lam},
\end{equation}
where the first term in the RHS of Eq. (\ref{lam}) is the number
of NN connected  to both extrems of the $k$-mer and the second
term is due to the NN connected to the $k-2$ inner units of the
molecule. The probability that one of the NN is filled by
molecules is equal to $kn/M$ (random distribution of molecules
among sites). Then, the mean number of pairs of filled sites is
given by
\begin{eqnarray}
\overline{E_k}(n,M) &    =   & \frac{1}{2} n\left[
2(\zeta-1)+(k-2)(\zeta-2)\right]
\left( \frac{kn}{M} \right) \nonumber \\
       &    =   & \frac{z k n^2}{2M}.
\end{eqnarray}

Then, the grand partition function can be written as,
\begin{equation}
\Xi=\sum_{n=0}^{n_{max}} \left[\eta(\zeta,k) \ \xi \right]^{n}
\frac{[M-(k-1)n]!}{n![M-kn]!} \exp{\left(- \frac{\beta w z k
n^2}{2M}  \right)} \label{xif}.
\end{equation}

When $w=0$, the summation in Eq. (\ref{xif}) can be performed very
easily. In the more general case $w \neq 0$, the summation cannot
be done in an easy way, but we can apply the method of the maximum
term. Then, the sum in Eq. (\ref{xif}) is replaced by its maximum
term, found from the condition $\frac{\partial \ln t}{\partial n}
= 0$, being $t(n,M,T)=\left[\eta \ \xi \right]^{n}
\frac{[M-(k-1)n]!}{n![M-kn]!} \exp{\left(- \frac{\beta w z k
n^2}{2M}  \right)}$. Thus,
\begin{equation}
\ln\eta \  \xi - \frac{\beta w z k n_1}{M}-(k-1)\ln
[M-(k-1)n_1]-\ln n_1 +k \ln (M-kn_1) = 0 \label{tm},
\end{equation}
where $n_1$ is the value of $n$ giving the maximum term in the sum
in Eq. (\ref{xif}).

Introducing $n_1$ in Eq. (\ref{xif}) and applying Stirling's
approximation, $\ln \Xi$ is given by
\begin{eqnarray}
\ln \Xi & = & n_1 \ln\eta \ \xi - \frac{\beta w z k
n_1^2}{2M}+[M-(k-1)n_1]\ln [M-(k-1)n_1] \nonumber \\
& & - n_1 \ln n_1 -(M-kn_1) \ln (M-kn_1) \label{lnxi}.
\end{eqnarray}

The average number of the molecules in the adsorption system $N$
is
\begin{equation}
N=k_B T \left(\frac{\partial \ln \Xi}{\partial \mu}\right)_{T,M}
\label{n}.
\end{equation}
Then,
\begin{eqnarray}
\left(\frac{\partial \ln \Xi}{\partial \mu}\right)_{T,M} & = &
n'_1 \ln\eta \ \xi + n_1 \frac{\xi' }{\xi} - \frac{\beta w z k n_1
n'_1}{M} -(k-1) n'_1 \ln [M-(k-1)n_1] \nonumber \\
& & - n'_1 \ln n_1 + k n'_1 \ln (M-kn_1) \label{dxidmu},
\end{eqnarray}
where $n'_1=\frac{\partial n_1}{\partial \mu}$ and $\xi'$ has the
following explicit form:
\begin{equation}
\xi'=\frac{\partial \xi}{\partial \mu} = \frac{1}{k_B T} \frac{q_1
\exp(\mu/k_{B}T)}{\left[1-q \exp(\mu/k_{B}T) \right]^2}
\label{xiprima}.
\end{equation}

Inserting the condition Eq. (\ref{tm}) into Eq. (\ref{dxidmu}), we
obtain
\begin{eqnarray}
\left(\frac{\partial \ln \Xi}{\partial \mu}\right)_{T,M} & = & n_1
\frac{\xi' }{\xi}  \label{dxidmuf}.
\end{eqnarray}

Considering now Eqs. (\ref{tm}), (\ref{n}), (\ref{xiprima}) and
(\ref{dxidmuf}), the adsorption isotherm equation can be obtained.
In the case of adsorbed monomers ($k=1,\eta=1$), Eq. (\ref{tm})
can be written as
\begin{equation}
\ln \xi - \frac{\beta w z n_1}{M} -\ln n_1 + \ln (M-n_1) = 0,
\label{tmk1}
\end{equation}
and
\begin{equation}
\theta_1=\frac{n_1}{M} = \frac{\xi \exp(-\beta w z
\theta_1)}{1+\xi \exp(-\beta w z \theta_1) }, \label{tmonk1}
\end{equation}
where $\theta_1=k n_1 / M$ is the monolayer coverage, being
$\theta= k N / M$ the total coverage. Then, from Eqs. (\ref{xi1}),
(\ref{n}), (\ref{xiprima}) and (\ref{tmonk1})
\begin{eqnarray}
\theta = \frac{N}{M} & = & k_B T \left(\frac{\partial \ln
\Xi}{\partial \mu}\right)_{T,M} \nonumber \\
& = & \left[\frac{1}{1-q \exp(\mu/k_BT)}\right] \nonumber \\
 & & \left[\frac{q_1 \exp(\mu/k_BT-w z \theta_1 /k_B T)}{1-q
\exp(-\mu/k_BT)+q_1 \exp(\mu/k_BT-w z \theta_1 /k_B T) }\right]
\label{betk1},
\end{eqnarray}
which can be easily recognized as the classical BET equation if we
write it in the form
\begin{eqnarray}
\theta & = & \frac{1}{\left(1-x\right)}
\frac{c^*x}{\left(1-x+c^*x\right)} \label{bet},
\end{eqnarray}
where $c^*= (q_1/q)\exp(-w z \theta_1/k_BT)$.

For the case of dimers ($k=2$), Eq. (\ref{tm}) reduces to
\begin{equation}
\ln\eta \ \xi - \frac{\beta w z n_1}{M} -\ln n_1 - \ln (M-n_1)
+\ln (M-2n_1)^2= 0, \label{tmk2}
\end{equation}
and
\begin{equation}
\theta_1=\frac{2n_1}{M} = 1-\frac{1}{\left[ 1+4\eta \ \xi
\exp(-\beta w z \theta_1) \right]^{1/2}}. \label{tmonk2}
\end{equation}

Using Eqs. (\ref{xi1}), (\ref{n}), (\ref{xiprima}) and
(\ref{tmonk2}) we obtain
\begin{eqnarray}
\theta = \frac{2N}{M} & = & 2k_B T \left(\frac{\partial \ln
\Xi}{\partial \mu}\right)_{T,M} \nonumber \\
& = & \frac{1}{1-q \exp(\mu/k_BT)} \left\{ 1-\frac{1}{\left[
1+4\eta \ \xi \exp(-\beta w z \theta_1) \right]^{1/2}}  \right\},
\label{betk2}
\end{eqnarray}
and, in terms of $c^*=\eta (q_1/q)\exp(-w z \theta_1/k_BT)$ and
$x$,
\begin{eqnarray}
\theta & = & \frac{1}{\left(1-x\right)} \left\{ 1-
\left[\frac{1-x}{\left(1+4c^*x-x\right)}\right]^{1/2} \right\}.
\label{bet2}
\end{eqnarray}
Eq. (\ref{bet2}) is similar to the recently reported multilayer
isotherm for non-interacting dimers \cite{SS10}. In this case,
taking into account the lateral interactions between the primarily
adsorbed molecules adds to the adsorption energy the term $\beta w
z \theta_1$, which represents the potential of the average force
acting on an admolecule in the first adsorbed layer from its NN in
the first layer.

For $k>2$ the explicit expression of the adsorption isotherm
cannot be obtained in a easy way. However, the calculations for
large molecules can be easily done through a standard computing
procedure; in our case, we used Maple software.

An alternative method to calculate the multilayer adsorption
isotherm was recently reported in Ref.\cite{SS10}. The theoretical
procedure can be described as follows:
\begin{itemize}
\item[1)] By using $\theta_1$ as a parameter ($0 \leq \theta_1
\leq 1$), the relative pressure is obtained by using the
condition
\begin{eqnarray}
{x} = {1 \over 1+c \lambda_{1}^{-1}}, \label{landamon}
\end{eqnarray}
where $\lambda_{1}$ is the monolayer fugacity. This calculation
requires the knowledge of an analytical expression for the
monolayer adsorption isotherm, $\lambda_{1} (\theta_{1})$.

\item[2)] The values of $\theta_{1}$ and $x$ are introduced in
\begin{equation}
\theta=\frac{\theta_1}{1-x}, \label{rel1n}
\end{equation}
and the total coverage is obtained.
\end{itemize}

The equivalence between both methodologies can be easily
understood. In fact, Eq. (\ref{landamon}) can be obtained from Eq.
(\ref{xi1}) and the maximum term condition Eq. (\ref{tm}). On the
other hand, from Eqs. (\ref{n}), (\ref{xiprima}) and
(\ref{dxidmuf}) and after some algebra, the total coverage can be
written in terms of the monolayer coverage, and Eq. (\ref{rel1n})
is recovered. As an example, in the following we show the use of
the method in Ref.\cite{SS10} to calculate the adsorption
isotherms in Eqs. (\ref{bet}) and (\ref{bet2}).

We start from the equation
\begin{equation}
\lambda_1={\theta_1 \over \eta(\zeta,k) \ k} {{\left[1-{{(k-1)}
\over {k}}\theta_1 \right]^{k-1}} \over {\left( 1-\theta_1
\right)^k}} \exp{\left( \beta z w \theta_1\right)} \label{landa},
\end{equation}
which represents the BWA isotherm of interacting $k$-mers adsorbed
at monolayer \cite{LANG5,SS12}.

Substituting Eq. (\ref{landa}) into Eq. (\ref{landamon}), one
obtains the following expression for the relative pressure,

\begin{equation}
{p \over {p_o}}={{\theta_1\left[1-{{(k-1)} \over {k}}\theta_1
\right]^{k-1} \exp{\left( \beta z w \theta_1\right)}} \over {c
\eta(\zeta,k) k \left( 1-\theta_1 \right)^k +
\theta_1\left[1-{{(k-1)} \over {k}}\theta_1 \right]^{k-1}
\exp{\left( \beta z w \theta_1\right)} }}. \label{ponpo1d}
\end{equation}

Eqs. (\ref{rel1n}) and (\ref{ponpo1d}) represent the mean-field
solution describing the adsorption of interacting $k$-mers at
multilayer regime on a homogeneous surface. In the case of monomer
[dimer] adsorption, Eqs. (\ref{rel1n}) and (\ref{ponpo1d}) reduce
to Eq. (\ref{bet}) [(\ref{bet2})].

We now turn to the QCA, which is significantly better than the
BWA. The important assumption in this method is that pairs of NN
sites are treated as if they were independent of each other (this
assumption is, of course, not true, because the pairs overlap
\cite{Hill}).

In order to apply the scheme in Ref. \cite{SS10}, we start with
the monolayer adsorption isotherm of interacting $k$-mers adsorbed
on a lattice of connectivity $\zeta$ obtained from the formalism
of QCA \cite{SS12},
\begin{equation}
\lambda_1=\left[\frac{\theta_1 \exp {\left( \beta w z /2
\right)}}{k \ \eta(\zeta,k) \left( \frac{2}{\zeta}
\right)^{2(k-1)}}\right] \left[
\frac{(1-\theta_1)^{k(\zeta-1)}\left[ k - (k-1) \theta_1
\right]^{k-1}
 \left[\frac{z \theta_1}{2k}- \alpha \right]^{z/2}}{\left[\frac{\zeta k}{2}-(k-1)\theta_1) \right]^{k-1} \left[\frac{\zeta}{2}(1-\theta_1)- \alpha \right]^{k \zeta /2} \left(\frac{z \theta_1}{\zeta k}
 \right)^{z}}\right],
 \label{mu}
\end{equation}
where  $\alpha$ is
\begin{equation}
\alpha = \frac{z \zeta}{2k} \frac{\theta_1
(1-\theta_1)}{\left[\frac{\zeta}{2}-\left(\frac{k-1}{k}
\right)\theta_1 + b\right]},    \label{alfa}
\end{equation}
\begin{equation}
b= \left\{ \left[\frac{\zeta}{2}-\left(\frac{k-1}{k}
\right)\theta_1 \right]^2  - \frac{z \zeta}{k} A \theta_1
(1-\theta_1) \right\}^{1/2},    \label{b}
\end{equation}
and
\begin{equation}
A= 1-\exp(-\beta w).
\end{equation}

Replacing Eq. (\ref{mu}) into Eq. (\ref{landamon}), we obtain
\begin{equation}
\left({p \over {p_o}}\right)^{-1}  =  1+\frac{c k \eta(\zeta,k)
\left( \frac{2}{\zeta} \right)^{2(k-1)} \left[\frac{\zeta
k}{2}-(k-1)\theta_1) \right]^{k-1}
\left[\frac{\zeta}{2}(1-\theta_1)- \alpha \right]^{k \zeta /2}
\left(\frac{z \theta_1}{\zeta k} \right)^{z}} {\theta_1 \exp
{\left( \beta w z /2 \right)}(1-\theta_1)^{k(\zeta-1)}\left[ k -
(k-1) \theta_1 \right]^{k-1}
 \left[\frac{z \theta_1}{2k}- \alpha \right]^{z/2}}
\label{pqca}.
\end{equation}
Eqs. (\ref{rel1n}) and (\ref{pqca}) represent the solution
describing the multilayer adsorption of interacting $k$-mers on
homogeneous surfaces in the framework of the QCA.

 \subsection{Monte Carlo Simulation of Adsorption in the Grand Canonical Ensemble}

The adsorption process is simulated through a grand canonical
ensemble Monte Carlo (GCEMC) method\cite{SS10}.

For a given value of the temperature $T$ and chemical potential
$\mu$, an initial configuration with $N$ $k$-mers adsorbed at
random positions (on $kN$ sites) is generated. Then, an
adsorption-desorption process is started, where each elementary
step is attempted with a probability given by the Metropolis
\cite{Metropolis} rule:
\begin{equation}
W =\min \left\{ 1, \exp{\left[- \beta \left( \Delta H - \mu \Delta
N  \right) \right]}\right\},
\end{equation}
where $\Delta H$ and $\Delta N$ represent the difference between
the Hamiltonians and the variation in the number of particles,
respectively, when the system changes from an initial state to a
final state. In the process there are four elementary ways to
perform a change of the system state, namely, adsorbing one
molecule onto the surface, desorbing one molecule from the
surface, adsorbing one molecule in the bulk liquid phase and
desorbing one molecule from the bulk liquid phase. In all cases,
$\Delta N= \pm 1$.

The algorithm to carry out one MC step (MCS), is the following :

\begin{itemize}
\item[1)]  Set the value of the chemical potential $\mu$ and the
temperature $T$.

\item[2)]  Set an initial state by adsorbing $N$ molecules in the
system. Each $k$-mer can adsorb in two different ways: $i)$ on a
linear array of ($k$) empty sites on the surface or $ii)$ exactly
onto an already adsorbed $k$-mer.

\item[3)]  Introduce an array, denoted as $A$, storing the
coordinates of $n_e$ entities, being $n_e$,
\begin{eqnarray}
n_e & = & {\rm number \ of \ available \ adsorbed \ {\it k}\rm{-mers} \ for \ desorption \ (n_d)} \nonumber \\
& + & {\rm \ number \ of \ available \ {\it k}\rm{-uples} \ for \
adsorption \ (n_a)},
\end{eqnarray}
where $n_a$ is the sum of two terms: $i)$ the number of $k$-uples
of empty sites on the surface and $ii)$ the number of columns of
adsorbed $k$-mers\footnote{Note that the top of each column is an
available $k$-uple for the adsorption of one $k$-mer.}.

\item[4)] Choose randomly one of the $n_e$ entities, and generate
a random number $\xi {\in }\left[ 0,1\right] $

\begin{itemize}
\item[4.1)]  if the selected entity is a $k$-uple of empty sites
on the surface then adsorb a $k$-mer if $\xi \leq W_{ads}^{surf}$,
being $W_{ads}^{surf}$ the transition probability of adsorbing one
molecule onto the surface.

\item[4.2)]  if the selected entity is a $k$-uple of empty sites
on the top of a column of height $i$, then adsorb a new $k$-mer in
the $i+1$ layer if $\xi \leq W_{ads}^{bulk}$, being
$W_{ads}^{bulk}$ the transition probability of adsorbing one
molecule in the bulk liquid phase.

\item[4.3)]  if the selected entity is a $k$-mer on the surface
then desorb the $k$-mer if $\xi \leq W_{des}^{surf}$, being
$W_{des}^{surf}$ the transition probability of desorbing one
molecule from the surface.

\item[4.4)]  if the selected entity is a $k$-mer on the top of a
column then desorb the $k$-mer if $\xi \leq W_{des}^{bulk}$, being
$W_{des}^{bulk}$ the transition probability of desorbing one
molecule from the bulk liquid phase.
\end{itemize}

\item[5)] If an adsorption (desorption) is accepted in $4)$, then,
the array $A$ is updated.

\item[6)]  Repeat from step $4)$ $M$ times.
\end{itemize}

In the present case, the equilibrium state could be well
reproduced after discarding the first $m\approx 10^6MCS$. Then,
averages were taken over $m^{\prime }\approx 10^6MCS$ successive
configurations. The total coverage was obtained as simple
averages,
\begin{equation}
\theta = {k\left< N \right> \over M},
\end{equation}
where $\left< N \right>$ is the mean number of adsorbed particles,
and $\left<...\right>$ means the time average over the MC
simulation runs.

The computational simulations have been developed for 1D chains of
$10^4$ sites, and square $L \times L$ lattices, with $L = 100$,
and periodic boundary conditions. With this lattice sizes we
verified that finite-size effects are negligible.

\section{Results}

In the present section, we will analyze the main characteristics
of the thermodynamic functions given in Subsection 2.2, in
comparison with simulation results for a lattice-gas of
interacting $k$-mers on 1D and 2D substrates.

\subsection{Exact solution in the 1D case}

In Figs. 1-3 we address the comparison between the analytical
adsorption isotherms for 1D substrates and MC simulations.
Different values of the parameter $c$, the lateral interactions
and the $k$-mer sizes  have been considered.

We start analyzing the case of $k=1$, $c=1$ and different values
of $w/k_BT=-2,-1,0,1$ and $2$ [see Fig. 1 (a)]. The case $w=0$
(standard BET model) has been widely discussed in the literature
(see, for instance, Ref. \cite{Gregg}) and it has been shown that
a shape of a Type II isotherm is obtained so long as $c$ exceeds
2. In the case of this figure, $c=1$ and the curve for
non-interacting particles has the general shape of a Type III
isotherm. For repulsive couplings, the interactions do not favor
the adsorption on the first layer and the isotherms shift to
higher values of pressure. On the other hand, attractive lateral
interactions facilitate the formation of the monolayer.
Consequently, the isotherms shift to lower values of $p/p_o$ and
their slope increases as the ratio $|w|/k_BT$ increases. With
respect to the shape of the curves, there exists a range of
$w/k_BT$ where the isotherms keep the shape of a Type III isotherm
(in this case, $-1 \leq w/k_BT \leq 1$). However, a knee appears
in the isotherms (the curves adopt the shape of a Type II
isotherm) as the ratio $|w|/k_BT$ increases. The shape of the knee
depends on the value of $|w|/k_BT$, becoming sharper as the value
of $w/k_BT$ becomes more negative. This point will be illustrated
more clearly in Fig. 5.

The effect of the $k$-mer size on the adsorption isotherm can be
understood by analyzing Figs. 1 (b) and 1 (c), where the study in
Fig. 1 (a) is repeated for $k=2$ and $k=4$, respectively. Two main
conclusions can be drawn from the figures. Namely, $1)$ the
difference between the curves corresponding to different values of
$w/k_BT$ and $2)$ the range of $w/k_BT$ where the isotherms do not
develop an inflection point diminish as $k$ is increased.

Figs. 2 and 3 show the effect of the parameter $c$ on the
adsorption isotherms. As can be observed, all curves exhibit a
pronounced knee as the parameter $c$ is increased. This effect can
be better visualized in the insets of Figs. 2 and 3, where a zoom
of the region of low pressure is presented. In the case of
attractive interactions, the knee appears around $\theta=1$ and
can be associated to the formation of the monolayer. In the case
of repulsive interactions, $k$-mers avoiding configurations with
NN heads arrange in a structure of alternating particles separated
by an empty site. Thus, for a given value of $k$ and strong
repulsive interactions, a marked knee is found at $\theta=k/(k +
1)$.

To complete the discussion of Figs. 1-3, we evaluate the reaches
and limitations of the two theoretical approximations studied. QCA
leads to exact solution in 1D systems. Consequently, MC
simulations in the grand canonical ensemble (symbols) fully agree
with the predictions from QCA (solid lines), which reinforces the
robustness of the two methodologies employed here. With respect to
BWA, two different behaviors are observed : $(i)$ for small values
of $|w|/k_BT$, the theoretical curves show a good agreement with
the simulation data and $(ii)$ for $|w|/k_BT>2$, appreciable
differences are observed between BWA and MC results. For strong
attractive couplings [see, for instance, the case corresponding to
$w/k_BT=-2$ in Fig. 1 (a)], a characteristic van der Waals loop is
observed in the adsorption isotherm and BWA incorrectly predicts a
phase transition for $\zeta = 2$. The shape of the isotherms is
fairly independent of the size of the molecules. However, the
disagreement between the BWA curves and the exact results turns
out to be significantly large for larger $k$-mer sizes [see, for
instance, Fig. 1 (c)].

The analysis of the curves in Figs. 1-3 indicates that the
appearance of a inflection point in the curves depends on $k$, $c$
and $w/k_BT$. This will be studied in detail in the following. The
point of inflection can be obtained in three steps: $(1)$
differentiating twice the adsorption isotherm equation to obtain
$d^2 \theta/dY^2$ (being $Y = p/p_o$ for the sake of simplicity);
$(2)$ equating the resulting expression to zero and solving for
$Y$ gives $Y_F$, the value of $p/p_o$ at the point of inflection;
and $(3)$ inserting $Y_F$ in the adsorption isotherm equation
gives $\theta_F$, the value of $\theta$ at the point of
inflection.

The location of the point of inflection ($X_F \equiv
\theta_F,Y_F$) is plotted in Fig. 4 for different values of $k$
and $w/k_BT$. The information is organized as follows: $(i)$ as in
Figs. 1-3, parts (a), (b) and (c) correspond to $k=1$, $k=2$ and
$k=4$, respectively; $(ii)$ the curves in (a), (b) and (c) were
obtained for different values of $w/k_BT$ (as indicated in the
caption of each figure); and $(iii)$ each point on a given curve
corresponds to a determined value of $c$.

In order to understand the basic phenomenology, we consider in the
first place the case corresponding to $w/k_BT=0$ (highlighted
curve). Clearly, the value of $\theta$ at the point of inflection
may deviate considerably from unity. However, there exist a
certain value of $c = c_m$, where the point of inflection
coincides with the point corresponding to the monolayer capacity.
Fig. 5 shows the values of $c_m$, obtained numerically, as a
function of $w/k_BT$ for four $k$-mer sizes ($k=1,2,4,10$). Two
regimes can be clearly differentiated according to the sign of the
lateral interactions. For attractive interactions, $c_m$ is not
defined in all the range of $w/k_BT$. Thus, for each $k$-mer size,
there exists a limit value of the lateral interaction,
$w_{min}/k_BT$,  below of which the coordinate $X_F$
characterizing the inflection point in the adsorption isotherm is
larger than one. In other words, the $X_F-Y_F$ diagrams
corresponding to values of  $w/k_BT$ below $w_{min}/k_BT$ do not
cross the line corresponding to $X_F=1$ (dashed line in Fig. 4).
The values of  $w_{min}/k_BT$ for $k=1,2,4,10$ are collected in
Table I. Finally, $c_m$ increases monotonically as the interaction
energy is increased in the range $w_{min}/k_BT< w/k_BT<0$. On the
other hand, $c_m$ shows an exponential dependence [$\exp(\kappa
w/k_BT)$] in the range $w/k_BT>0$, where $\kappa$ is a parameter
depending on the $k$-mer size. The different values of $\kappa$,
obtained from the slope of $\ln c_m$ vs. $w/k_BT$ are reported in
Table I.

For values of $c$ between $c_m$ and infinity the adsorption at the
point of inflection exceeds the monolayer capacity; for values of
$c$ below $c_m$ the two quantities deviate more and more and for a
limit value of $c = c_n$, the point of inflection
disappears\footnote{For $c > c_n$, the isotherm is of Type II and
when $c$ is less than $c_n$ the isotherm is of Type III and
discussion of the point of inflection is meaningless.}. In the
low-coverage regime ($\theta \rightarrow 0$), $c_n$ vs. $w/k_BT$
can be calculated analytically (see Appendix for further
discussion). The result of this calculation is presented in Fig. 6
for different values of $k$. Solid lines represent theoretical
data from Eq. (\ref{fin})  and symbols correspond to values of
$c_n$ obtained numerically. As can be visualized from the figure,
the dashed line separates two well differentiate regions. At right
of the dashed line, the coordinates of the inflection point
corresponding to the limit value $c_n$ are $X_F=0$ and $Y_F=0$.
Then, the assumption of $\theta \rightarrow 0$ in Appendix is
valid and the symbols coincide with the solid line. At left of the
dashed line, the point of inflection disappears for $X_F>0$ and
the solid line is not defined. The value of $w/k_BT$ corresponding
to the dashed line was obtained from the condition
$2k+1-2e^{-w/k_B T}=0$ [see Eq. (\ref{fin})]. From the point of
view of the $X_F-Y_F$ diagrams, the previous condition separates
diagrams defined in the origin $(X_F=0,Y_F=0)$ from those where
the inflection point disappears for $(X_F>0,Y_F>0)$ (see solid
circles in Fig. 4).

In the next we will refer to one of the main applications of BET
model, which consists in taking an experimental isotherm in the
low-pressure region and fitting values of the monolayer volume and
the parameter $c$, from the linearized form of the BET equation,
\begin{equation}
\frac{p/p_o}{[v(1- p/p_o)]} = \frac{1}{c v_m} + \frac{(c -
1)}{cv_m} p/p_o.
\end{equation}
The plot of $(p/p_o)/[v(1- p/p_o)]$ vs $p/p_o$ should therefore be
a straight line with slope $s=(c-1)/cv_m$ and intercept
$i=1/cv_m$. Solution of these two simultaneous equations gives
$v_m$ and $c$:
\begin{equation}
v_m = \frac{1}{s+i} \ \ {\rm and} \ \ c = \frac{s}{i}+1.
\end{equation}
In this context, it is of interest to study the behavior of
multilayer isotherms of $k$-mers (with $k > 2$) in the
low-pressure region\footnote{Although in each particular case it
is possible to find an optimum range of relative pressures, for
practical purposes, we have chosen to set this range from $0.05$
to $0.25$. Nevertheless, by choosing other ranges (for example,
between $0.05$ and $0.35$) we obtain similar results.} in
comparison with BET isotherm. For this purpose, we will analyze,
by using the standard BET formalism, exact theoretical isotherms
in the 1D case. As an example, Fig. 7 shows the results obtained
for $k = 2$, $c = 10$, $w/k_BT=-2,-1,0,1,2$  and pressures ranging
from $p/p_o = 0$ up to $p/p_o = 0.30$.  Symbols represent
theoretical data from QCA (exact results) and lines correspond to
linearized forms of the BET equation. A linear function is only
obtained if $k = 1$ and $w/k_BT=0$ (see inset). The nonlinear
behavior of interacting $k$-mers isotherms at low pressures, which
is a distinctive characteristic of many experimental isotherms, is
showing that the polyatomic character of the adsorbate and the
lateral interactions must be taken into account. The significant
differences observed as $k$ and $w/k_BT$ are varied indicate that
the analysis of experimental isotherms of interacting larger
molecules by means of the BET isotherm would lead to values of the
parameters $c$ and $v_m$ appreciably different from the real ones.

In order to measure the differences mentioned above, the analysis
of Fig. 7 was repeated in Fig. 8 for $c=10$ [part (a)], $c=100$
[part (b)] and different values of $k$ and $w/k_BT$. The results
obtained for the monolayer volume are shown in the figure, where
$v_m$ and $v_{BET}$ represent the real monolayer volume and the
corresponding value obtained from the BET fitting, respectively.
We start analyzing the case of $w/k_BT=0$ (see inset). In this
condition, the difference between $v_m$ and $v_{BET}$
($v_{BET}/v_m$) increases (decreases) initially as the $k$-mer
size is increased and remains almost constant for larger values of
$k$. As is shown in the figure, these differences diminish as the
parameter $c$ increases. Now, it is interesting to analyze the
effect of the lateral interactions. As was discussed above,
attractive lateral interactions favor the formation of the
monolayer and, consequently, compensate the effect of the
multisite occupancy. Thus, for a given value of $k$, $v_{BET}/v_m$
tends to one as $|w|/k_BT$ is increased. On the other hand,
repulsive interactions do not facilitate the formation of the
monolayer, increasing the differences between $v_m$ and $v_{BET}$.
For $w/k_BT>>1$, a marked knee is found at $\theta=k/(k + 1)$ and
$v_{BET}$ is close to this value of coverage. As the $k$-mer size
is increased, $\theta=k/(k + 1) \rightarrow 1$ and, consequently,
$v_{BET}/v_m \rightarrow 1$.  This phenomenon can be clearly
visualized by observing the curves corresponding to $w/k_BT=2$ in
the insets of Figs. 8 (a) and 8 (b).

\subsection{Approximate solution in the 2D case}

Because the structure of lattice space plays such a fundamental
role in determining the statistics of $k$-mers, it is of interest
and of value to inquire how a specific lattice structure
influences the main thermodynamic properties of adsorbed $k$-mers.
Following this line of thought, we use in this section the
lattice-gas language again and assume the same model as in
Subsection 3.1 with one exception: the sites form a 2D square
lattice instead of a 1D lattice.

In Fig. 9 simulated isotherms are compared to theoretical ones
from Eqs. (\ref{rel1n}), (\ref{ponpo1d}) and (\ref{pqca}) for
dimers adsorbed on 2D lattices with different values of $c$ and
$w/k_BT$: (a), $c=1$ and $w/k_BT=-1,-0.5,0,0.5,1,2$; (b) $c=10$
and $w/k_BT=-2,-1,0,1,2$; and (c) $c=100$ and
$w/k_BT=-2,-1,0,1,2$.

In the attractive case, the two theoretical approximations agree
qualitatively well and the adsorption isotherms for BWA (dashed
lines) and QCA (solid lines) are hardly distinguishable from each
other. The differences between numerical and theoretical results
can be much easily rationalized with the help of the absolute
error, $\Delta \theta(p/p_o)$, which is defined as $\Delta
\theta(p/p_o)=|{\theta_{theor}-\theta_{sim}}|_{p/p_o}/\theta_{sim}$,
where $\theta_{sim}$ ($\theta_{theor}$) represents the surface
coverage obtained by using MC simulation (analytical approach).
Each pair of values ($\theta_{sim}$, $\theta_{theor}$) is obtained
at fixed $p/p_o$.  The curves of $\Delta \theta(p/p_o)$ vs $p/p_o$
(data are not shown here for sake of simplicity) indicate that, in
all cases, QCA leads to appreciably better results than BWA.

Note that the stronger the lateral interaction, the more steep the
adsorption isotherm becomes. This behavior could be indicative of
the existence of a first-order phase transition at low
temperatures. However, the study of the critical behavior of the
system is out of the scope of the present work and will be object
of future studies.

With respect to repulsive interactions, the differences between
QCA and BWA are very appreciable. Beyond quantitative
discrepancies, there exists qualitative differences between both
approximations. Thus, while QCA is practically independent of $c$,
the discrepancies between BWA and MC results diminish appreciably
for larger values of $c$.

Summarizing, QCA gives a much better description of the 2D MC
adsorption isotherms than the BWA. In the particular case of
repulsive interactions, the disagreement between MC and BWA turns
out to be significantly large, while QCA appears as the simplest
approximation capable to take into account the main features of
the multisite-occupancy adsorption. In fact, there exists a wide
range of $w/k_BT$'s ($-2 \leq w/k_BT \leq 2$), where QCA provides
an excellent fitting of the simulation data. In addition, most of
the experiments in surface science are carried out in this range
of interaction energy. Then, QCA not only represents a qualitative
advance in the description of the multilayer adsorption of
$k$-mers with respect to the BWA, but also gives a framework and
compact equations to consistently interpret thermodynamic
adsorption experiments of polyatomics species such as alkanes,
alkenes, and other hydrocarbons on regular surfaces.

As indicated in the previous section, it is of interest to study
the behavior of 2D multilayer isotherms (with $k > 2$) in the
low-pressure region in comparison with BET isotherm. For this
purpose, the adsorption isotherms are plotted in the low-pressure
regime and fitted with the linearized form of the BET equation. A
typical example is shown in Fig. 10. The values of the parameters
used in the figure are: $k = 2$, $c = 10$ and $w/k_BT=-1,0,1$.
Open symbols represent Monte Carlo data, full symbols correspond
to theoretical results obtained from QCA and lines correspond to
linearized forms of the BET equation. As discussed in Fig. 9, QCA
provides very good results in the limit of attractive lateral
interactions and its accuracy diminishes for repulsive ad-ad
interactions.

The analysis of Fig. 10 was repeated for different values of $c$
and $w/k_BT$. In all cases, the monolayer volume was calculated
from the slope and intercept of the linearized form of the BET
equation. The results are shown in Fig. 11. Several conclusions
can be drawn from the figure: $(1)$ QCA agrees very well with the
numerical results in all range of $c$ and $w/k_BT$ studied; $(2)$
the differences between $v_m$ and $v_{BET}$ diminish as the
parameter $c$ increases; and $(3)$ attractive lateral interactions
compensate the effect of the multisite occupancy. In other words,
the behavior of $v_{BET}/v_m$ vs $w/k_BT$ is similar to the one
described above for the 1D case.

Finally, the effect of the $k$-mer size on $v_{BET}/v_m$ is
analyzed in Fig. 12. In the figure, simulation results for $k=2$
(full symbols) are compared with the corresponding ones obtained
for $k=4$ (crossed symbols). As can be observed, the behavior of
$v_{BET}/v_m$ vs $w/k_BT$ does not significantly vary as the
$k$-mer size changes from $k=2$ to $k=4$. In addition, the
agreement between QCA and MC data (not shown here for sake of
clarity) remains very good. Even though MC simulations of larger
linear adsorbates on regular 2D lattices would be necessary to
confirm the applicability of Eqs. (\ref{rel1n}) and (\ref{pqca}),
it should be pointed out that QCA is a good analytical approach
considering the complexity of the physical situation which is
intended to be described.

\section{Conclusions}

In this work, we have studied the multilayer adsorption of
interacting polyatomic molecules.  Two analytic isotherms were
developed in the framework of the BWA and the QCA.  The polyatomic
character of the adsorbate was modelled by a lattice-gas of
$k$-mers. With respect to lateral interactions, the ad-ad
couplings in the monolayer were explicitly considered in the
solutions. The range of validity of both isotherms was analyzed by
comparing theoretical and MC simulation results.

The new formalism from QCA leads to exact results in 1D and
provides a close approximation to study multilayer adsorption of
interacting polyatomics on 2D surfaces. On the other hand, the
artificial effects that the BWA induces on the main thermodynamic
functions can now be rationalized and compared with other
analytical approaches. In this sense, we have shown that the
disagreement between BWA and MC simulations increases as $(i)$ the
temperature is decreased (or the ratio $w/k_BT$ is increased) and
$(ii)$ the $k$-mer size is increased.

In addition, we have studied the 1D and 2D BET plots obtained
using the analytic and simulation isotherms. For non-interacting
$k$-mers, we found that the use of BET equation leads to an
underestimate of the true monolayer volume: this volume diminishes
as $k$ is increased. The situation is different for the case of
interacting molecules. Thus, attractive lateral interactions favor
the formation of the monolayer and, consequently, compensate the
effect of the multisite occupancy. In this case, the monolayer
volume predicted by BET equation agrees very well with the
corresponding true value. In the case of repulsive couplings, the
lateral interactions impede the formation of the monolayer and the
BET predictions are bad (even worse than those obtained in the
non-interacting case). Both the compensation effect for attractive
interactions and the underestimation of the monolayer volume for
repulsive interactions are more important for 2D systems.

\section{ACKNOWLEDGMENTS}

This work was supported in part by CONICET (Argentina) under
project PIP 6294; Universidad Nacional de San Luis (Argentina)
under project 322000; Universidad Tecnol\'ogica Nacional, Facultad
Regional San Rafael (Argentina) under project PID PQCO SR 563 and
the National Agency of Scientific and Technological Promotion
(Argentina) under project 33328 PICT 2005. The numerical work were
done using the BACO parallel cluster (composed by  60 PCs each
with a 3.0 GHz Pentium-4 processor) located  at Instituto de
F\'{\i}sica Aplicada, Universidad Nacional de San Luis-CONICET,
San Luis, Argentina.

\newpage

\section{Appendix 1}

In order to determine $c_n$, we will use a similar scheme to that
employed in Appendix A of Ref.\cite{SS10}. Here, we restrict the
analysis to 1D systems.

We start calculating the inflection point of the adsorption
isotherm
\begin{equation}
\frac{d^2\theta}{d^2(p/p_0)}=0    \ \ \ \ \ {\rm and} \ \ \ \ \
p/p_0 \rightarrow 0 \ \ {\rm (low \ density)}. \label{E1}
\end{equation}
By calculating the second derivative of $\theta$ in Eq.
(\ref{rel1n}), we obtain:
\begin{equation}
\theta^{''}=\frac{\theta_1^{''}}{(1-p/p_0)}+\frac{2\theta_1^{'}}{(1-p/p_0)^2}+\frac{2\theta_1}{(1-p/p_0)^3},
\label{E2}
\end{equation}
where $\theta^{''}$ and $\theta^{'}$ represent $\partial \theta /
\partial (p/p_o)$ and $\partial^2 \theta / \partial (p/p_o)^2$,
respectively. Now, by taking $\lim_{p/p_0 \rightarrow 0}$ in Eq.
(\ref{E2}), which implies $\lim_{\theta_1\rightarrow 0}$, the
following relation is obtained:
\begin{equation}
0=\theta_1^{''}+2\theta_1^{'}. \label{E3}
\end{equation}
The last equation allows us to obtain $c_n$ from the monolayer
adsorption isotherm. At low density ($\theta_1\rightarrow 0$), the
virial expansion approach can be considered as an exact result
\cite{LANG9}. As usual, $\lambda_{1}(\theta _1)$ can be written as
a power series. Thus, by using Eq. (\ref{landamon}),
\begin{equation}
p/p_0= \frac{\lambda_{1}}{\lambda_{1} + c}= \frac{\sum _{i=0}
^\infty a_i \theta_1 ^i}{\sum _{i=0} ^\infty a_i \theta_1 ^i + c}
\label{ecdepresion}.
\end{equation}

Differentiating both sides of the last equation with respect to
$p/p_0$, we obtain
\begin{equation}
1=\frac{\lambda_{1} ' (\lambda_{1} + c) + \lambda_{1} \lambda_{1}
'}{(\lambda_{1} + c )^2}= \frac{\lambda_{1} ' c}{(\lambda_{1} +
c)^2}
\end{equation}
\begin{equation}
1 = \frac{c \sum _{i=0} ^\infty i a_i \theta_1 ^{i-1} \theta _1
'}{\left(\sum _{i=0} ^\infty a_i \theta_1 ^i + c\right)^2}
\label{desrrollodelambda}
\end{equation}
Note that $\lambda_{1} \to 0$ as $\theta _1 \to 0$ and,
consequently, $a_0 = 0$. By taking the limit as $\theta _1 \to 0$,
Eq. (\ref{desrrollodelambda}) results
\begin{equation}
1 =  \frac{c_n a_1 \theta _1 '}{ \left( c_n \right)^2}
\end{equation}
and
\begin{equation}
\theta ' _1 = \frac{c_n}{a _1} \label{ec2}
\end{equation}

Now, by calculating the second derivative of Eq.
(\ref{ecdepresion}), we obtain
\begin{equation}
0=\frac{c\lambda_{1} '' (\lambda_{1}+c)^2 - c \lambda_{1} ^2
2(\lambda_{1} +c)}{(\lambda_{1} +c)^4}= \frac{c(\lambda_{1}
+c)}{(\lambda_{1} +c)^4}\left( \lambda_{1} '' \lambda_{1} +
\lambda_{1} '' c-2 \lambda_{1}  ^{'2} \right)
\end{equation}
\begin{equation}
0= \frac{\lambda_{1} '' c^2 - 2 \lambda_{1}  ^{'2} c}{c^3} = \frac
{\lambda_{1} '' c- 2 \lambda_{1} ^{'2} }{c^2}
\end{equation}
\begin{equation}
0= \lambda_{1} '' c- 2 \lambda_{1}  ^{' 2}   \label{ec_1}
\end{equation}
being
\begin{equation}
\lambda_{1} ' = \sum _{i=0} ^\infty a_i \theta _1 ^{i-1} i \theta
_1' \label{l}
\end{equation}
and
\begin{equation}
\lambda_{1} '' = \sum _{i=0} ^\infty a_i ( \theta _1 ^{i-2} i(i-1)
\theta _1'^2+\theta _1 ^{i-1} i \theta _1 ''). \label{ll}
\end{equation}

By taking the limit as $\theta _1 \to 0$, Eqs. (\ref{l}) and
(\ref{ll}) can be written as:

\begin{equation}
\lambda_{1} '(\theta _1 \to 0) =   a_1  \theta _1' \label{l0}
\end{equation}
\begin{equation}
\lambda_{1} '' (\theta _1 \to 0)=2  a_2 \theta _1'^2+ a_1 \theta
_1 ''.\label{ll0}
\end{equation}
Then, by introducing Eqs. (\ref{l0}) and (\ref{ll0}) in Eq.
(\ref{ec_1}) and by using Eq. (\ref{ec2}), we obtain
\begin{equation}
0=-2a_1^2 \frac{c_n^2}{a_1 ^2}+ c _n a_1 \theta _1 '' +2 c_n a_2
\frac{c _n^2}{a_1 ^2}
\end{equation}
\begin{equation}
0=-2  c_n^2+ c _n a_1 \theta _1 '' +2 c_n a_2 \frac{c _n^2}{a_1
^2},
\end{equation}
and
\begin{equation}
\theta _1 '' = \frac{2c_n ^2 - 2 \frac{a_2 c_n^2}{a_1 ^2}}{c_n
a_1}= \frac{2c_n}{a_1}-\frac{2 a_2 c_n^2}{a_1 ^3}.\label{con2}
\end{equation}

Finally, by introducing Eqs. (\ref{ec2}) and (\ref{con2}) in Eq.
(\ref{E3}), $c_n$ can be written in terms of $a_1$ and $a_2$:
\begin{equation}
0=\frac{2c _n}{a_1} - \frac{2 a_2 c_n^2}{a_1 ^3} + \frac{2
c_n}{a_1}
\end{equation}
and
\begin{equation} c_n=\frac{2 a_1 ^2}{a_2}. \label{eq_3}
\end{equation}

Now, we start with the calculation of $a_1$ and $a_2$
\cite{LANG9}. For this purpose, we write the first two terms of
the grand partition function of a lattice-gas of $N$ $k$-mers on
$M$ sites,
\begin{equation}
\Xi_k (M,\lambda_{1})= 1 + Q_k(M,1) \lambda_{1} + Q_k(M,2)
\lambda_{1}^2 \cdots
\end{equation}
where $Q_k(M,1)$ and $Q_k(M,2)$ represent the partition functions
for one and two $k$-mers, respectively.

By using the following expression,
\begin{eqnarray}
\ln\left[1+f(x) \right] & = & \ln\left[1+f(0) \right] + \left.
\left[\frac{f'(x)}{1+f(x)} \right] \right|_{x=0} x + \frac{1}{2!}
\left. \left\{ \frac{f''(x)}{1+f(x)} - \left[\frac{f'(x)}{1+f(x)}
\right]^2 \right\} \right|_{x=0} x^2 +
\nonumber\\
&  & \frac{1}{3!} \left. \left\{ \frac{f'''(x)}{1+f(x)} -3
\frac{f''(x)f'(x)}{\left[1+f(x) \right]^2} + 2
\left[\frac{f'(x)}{1+f(x)} \right]^3 \right\} \right|_{x=0} x^3 +
\cdots,
\end{eqnarray}
where $f(x)$ is an arbitrary function, $f'(x)=d f / dx$,
$f''(x)=d^2 f / dx^2$ and $f'''(x)=d^3 f / dx^3$, $\ln \Xi_k
(M,\lambda_{1})$ can be written as
\begin{equation}
\ln \Xi_k (M,\lambda_{1})= Q_k(M,1) \lambda_{1} + \frac{1}{2}
\left[2Q_k(M,2) - Q^2_k(M,1) \right] \lambda_{1}^2 + \cdots
\end{equation}
and the surface coverage results
\begin{eqnarray}
\theta  = k \frac{N}{M} & = & k \frac{\lambda_{1}}{M} \left.
\frac{\partial \ln \Xi_k}{\partial \lambda_{1}}\right|_{M,k}
\nonumber\\
& = & \frac{k}{M} Q_k(M,1) \lambda_{1} + \frac{k}{M}
\left[2Q_k(M,2) -
Q^2_k(M,1) \right] \lambda_{1}^2 + \cdots \nonumber\\
& = & b_1 \lambda_{1} + b_2 \lambda_{1} ^2 + \cdots = \sum _{i=0}
^{\infty} b_i \lambda_{1} ^i
\end{eqnarray}
where the $b_i$'s are the well-known virial coefficients. In this
case,
\begin{equation}
b_1= \frac{k}{M} Q_k (M,1)
\end{equation}
and
\begin{equation}
b_2= \frac{k}{M} \left[ 2 Q_k (M,2)- Q_k (M,1) ^2\right].
\end{equation}
In addition, the relationship between the $b_i$'s and the $a_i$'s
can be obtained by simple algebra. Thus,
\begin{equation}
a_1 = \frac{1}{b_1}
\end{equation}
and
\begin{equation}
a_2 = - \frac{b_2}{b_1 ^3}.
\end{equation}

On the other hand, $Q_k(M,1)$ and $Q_k(M,2)$ can be calculated as
\begin{equation}
Q_k (M,1)=M
\end{equation}
and
\begin{equation}
Q_k (M,2) = g_0 e^{- 0/k_B T} + g_1 e^{-w /k_B T}
\end{equation}
where the configurational factors can be easily obtained. Thus,
\begin{equation}
g_0 = \frac{1}{2 !} M \left[M-(2k+1) \right]
\end{equation}

\begin{equation}
g_1= \frac{1}{2!} M \ 2.
\end{equation}

Then,
\begin{equation}
b_1 = \frac{k}{M} \Omega _k (M,1) = k,
\end{equation}
\begin{equation}
a_1 = \frac{1}{b_1} = \frac{1}{k},
\end{equation}
\begin{eqnarray}
b_2 & = & \frac{k}{M} \left[ 2 Q_k (M,2)- Q_k (M,1) ^2\right]
\nonumber\\
& = & \frac{k}{M} \left\{ M \left[M- (2k+1)\right] + 2
M
e^{-w/k_BT} -M^2 \right\} \nonumber\\
& = & 2 k e^{-w/k_B T}- (2k+1) k,
\end{eqnarray}
\begin{equation}
a_ 2 =  - \frac{b_2}{{b_1}^3} = \frac{2k+1 - 2 e^{-w / k_B
T}}{k^2},
\end{equation}
and, finally,
\begin{eqnarray}
 c_n & = & \frac {2 a_1 ^2}{a_2} \nonumber\\
& = & \frac{2}{2k+1-2e^{-w/k_B T}} \label{fin}
\end{eqnarray}

In the case $w=0$, Eq. (\ref{fin}) reduces to the expression
obtained previously for noninteracting admolecules \cite{SS10}.

\newpage

\begin{figure}[th]
\centerline{\psfig{file=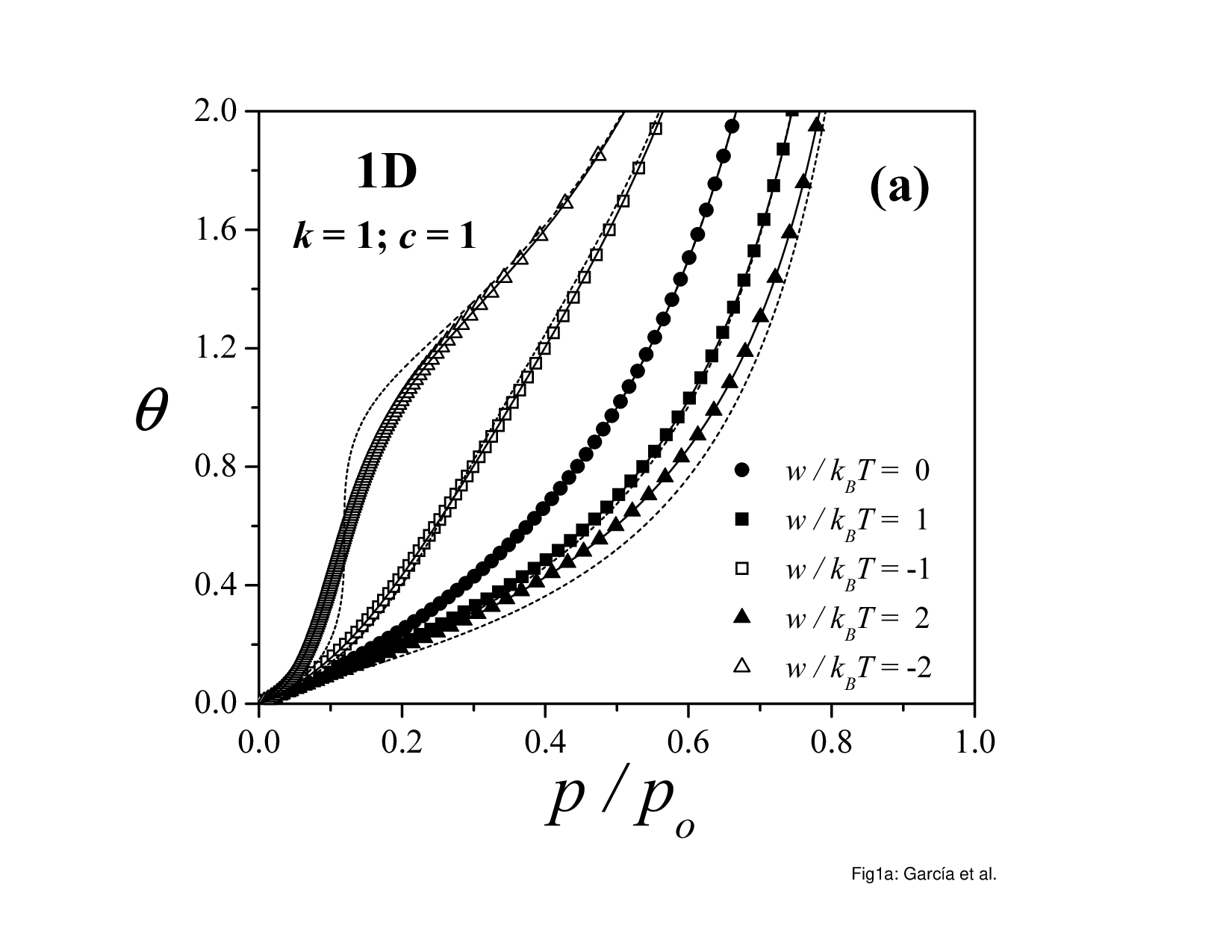,width=9cm}}
\centerline{\psfig{file=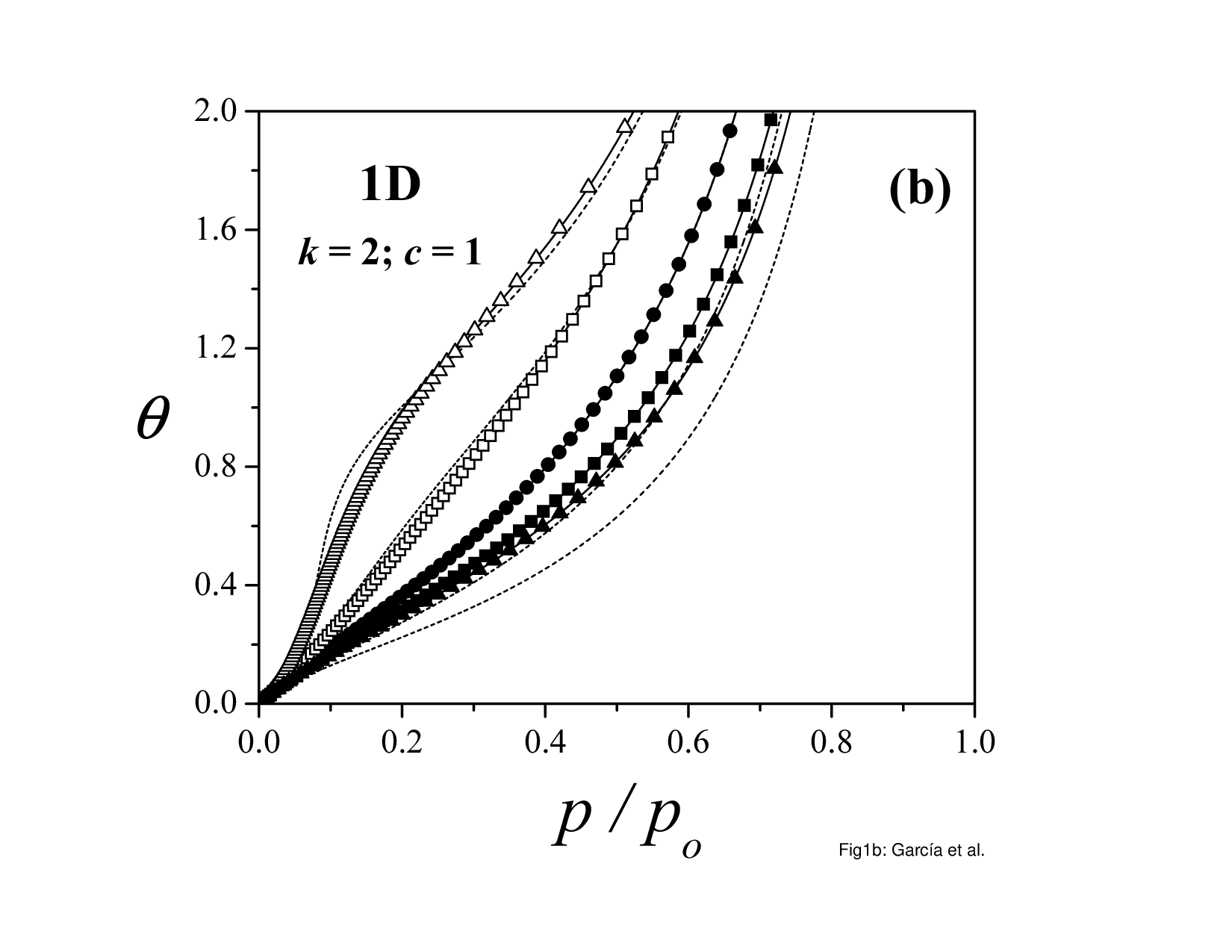,width=9cm}}
\centerline{\psfig{file=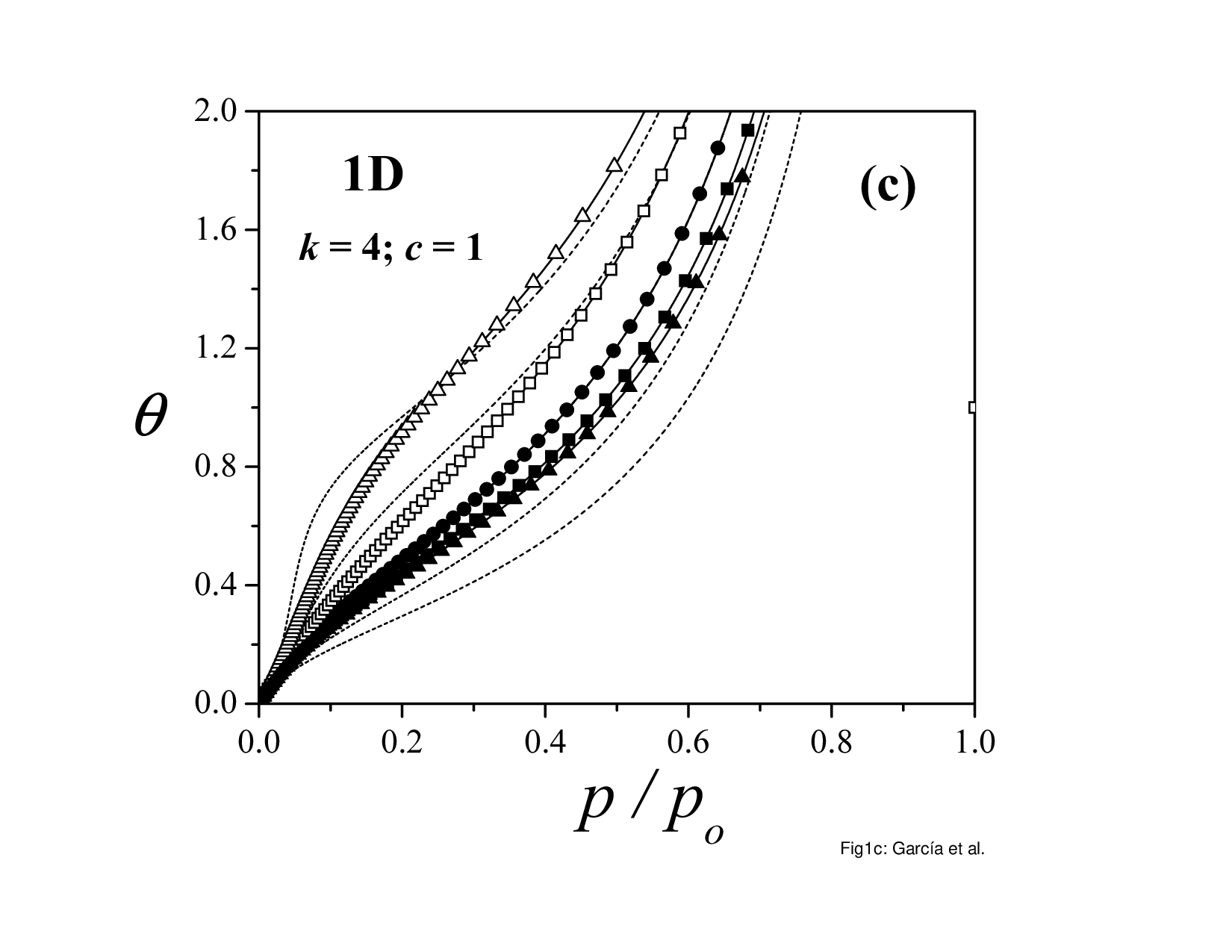,width=9cm}}
\caption{Adsorption isotherms for $k$-mers on 1D lattices, $c=1$
and different values of $w/k_BT$ (as indicated). (a) $k=1$; (b)
$k=2$ and (c) $k=4$. Symbols, solid lines and dashed lines
represent results from Monte Carlo simulations, QCA and BWA,
respectively.}
\end{figure}

\begin{figure}[th]
\centerline{\psfig{file=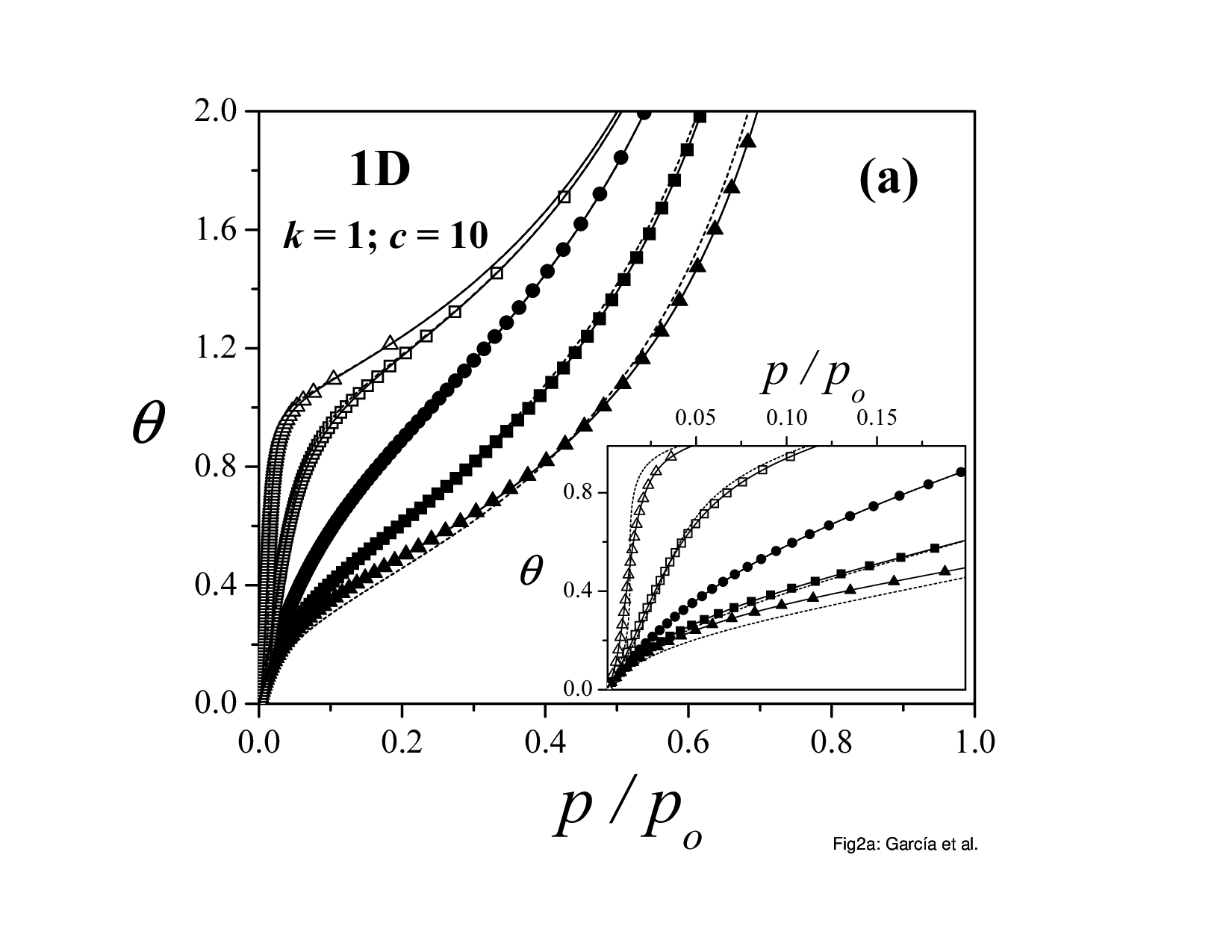,width=9cm}}
\centerline{\psfig{file=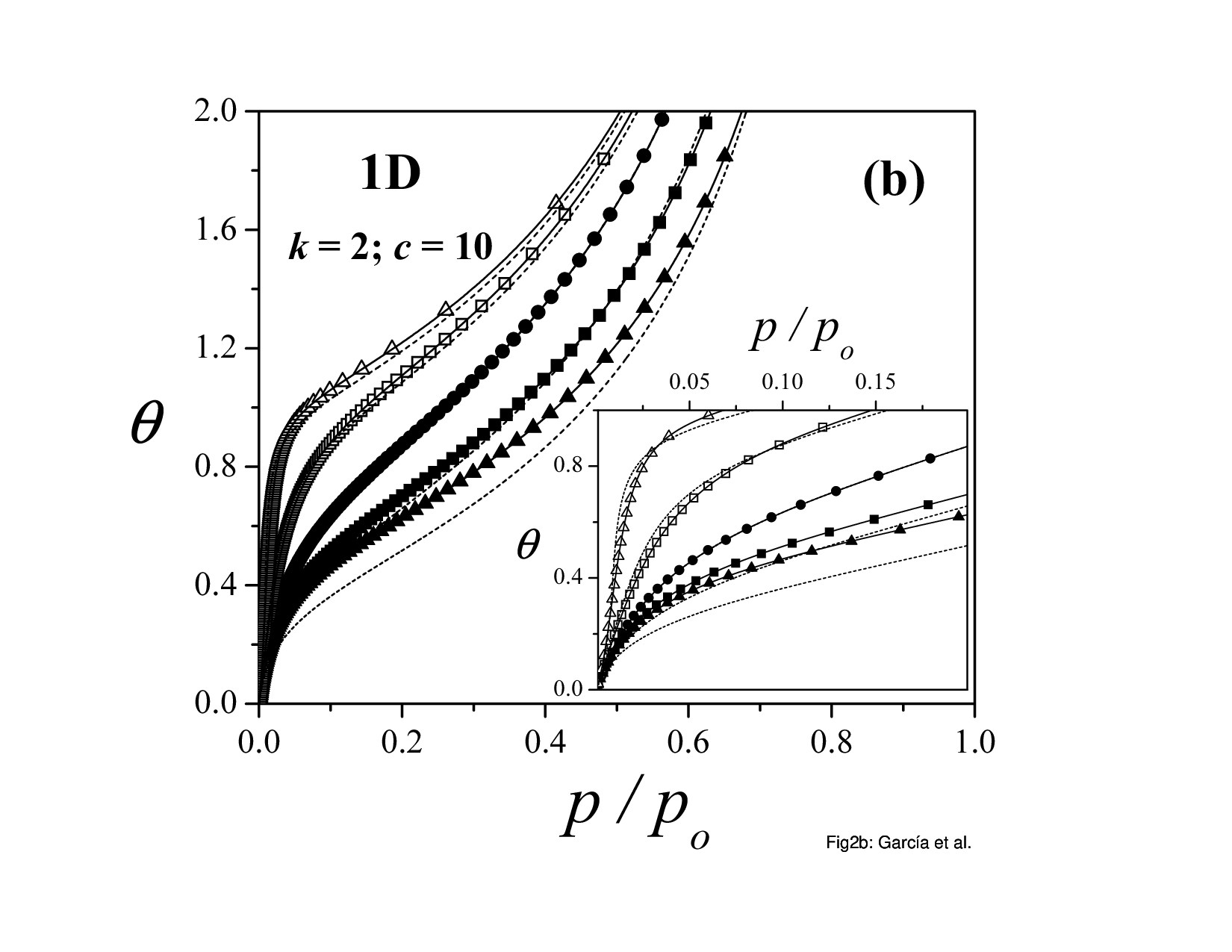,width=9cm}}
\centerline{\psfig{file=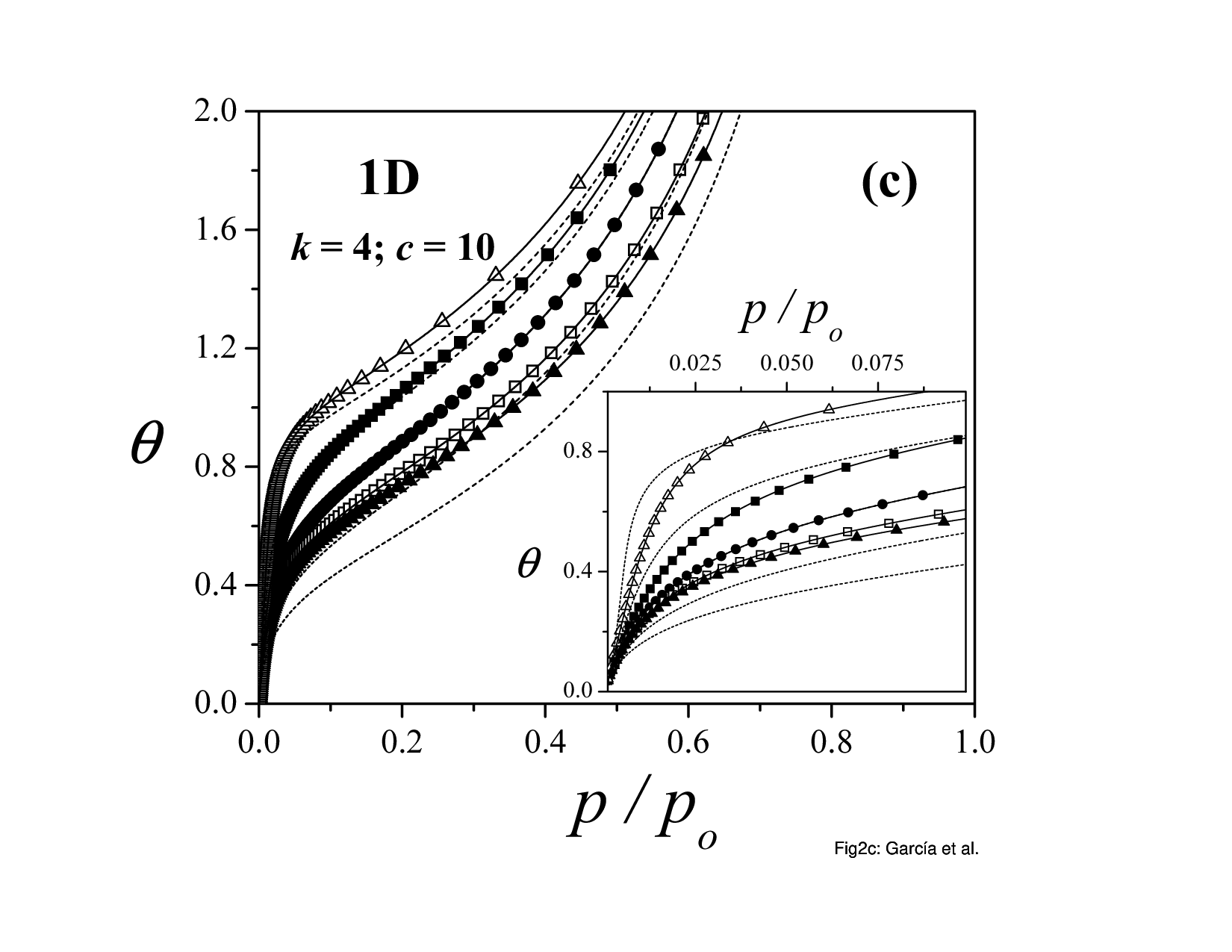,width=9cm}} \caption{As  Fig.
1 for $c=10$. A zoom of the low-pressure region is presented in
the inset of each figure.}
\end{figure}

\begin{figure}[th]
\centerline{\psfig{file=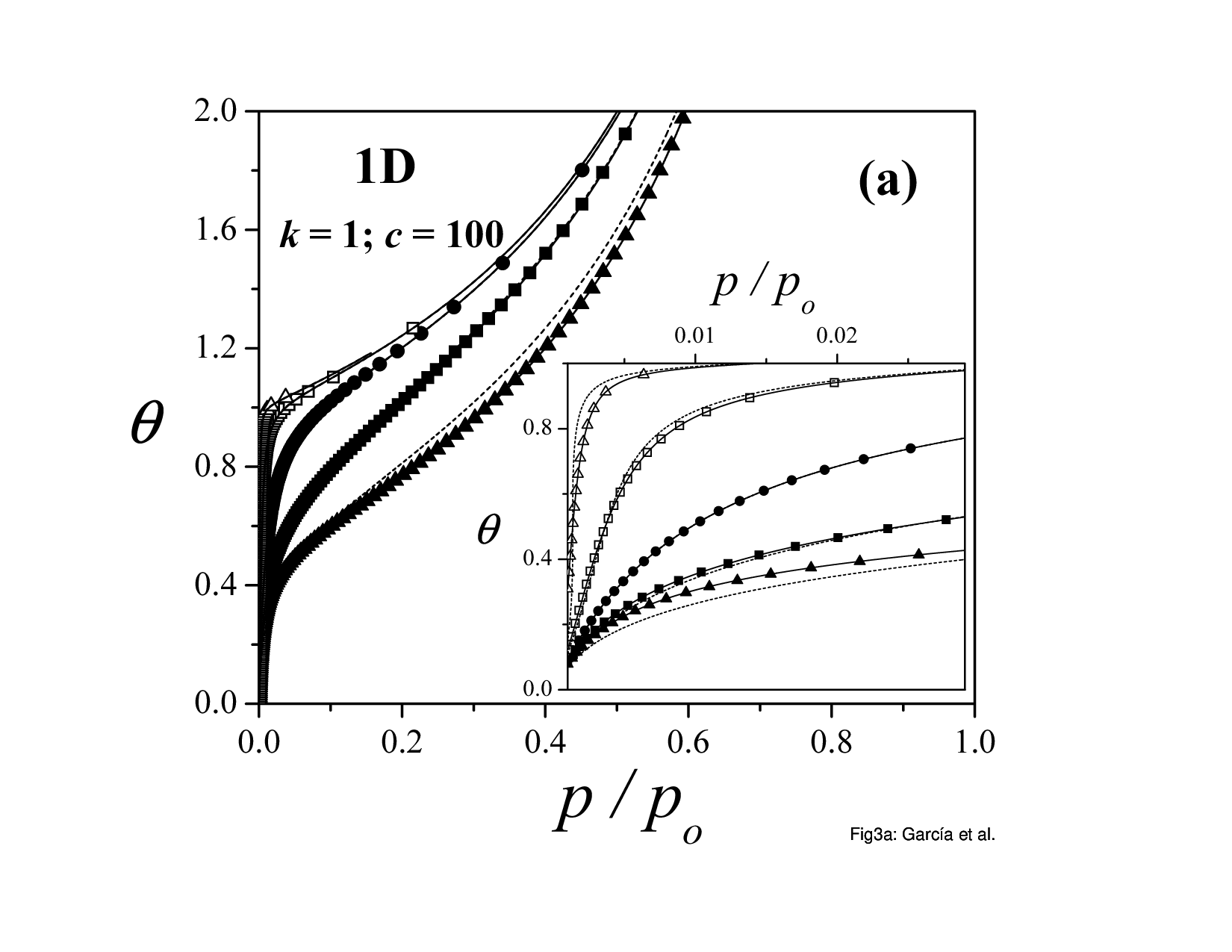,width=9cm}}
\centerline{\psfig{file=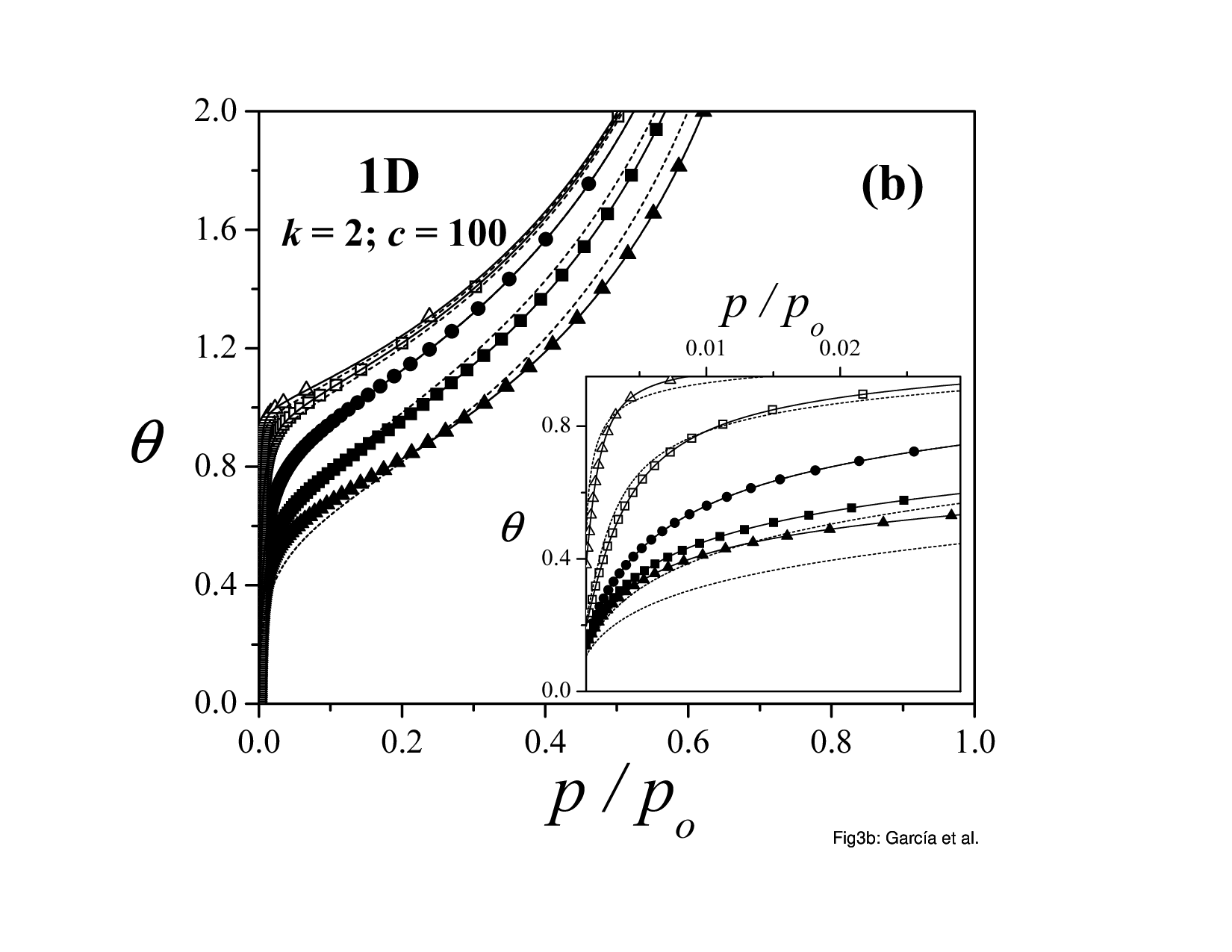,width=9cm}}
\centerline{\psfig{file=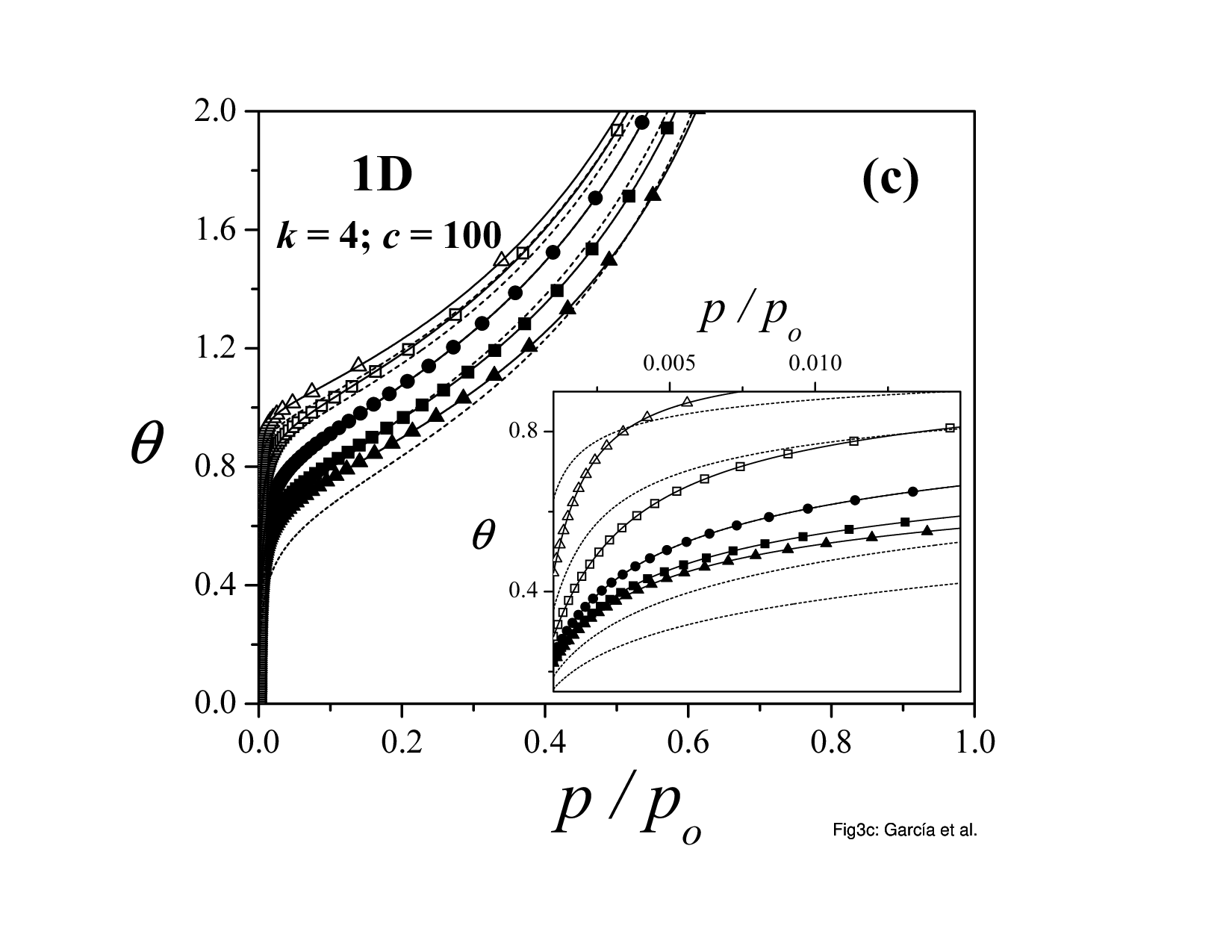,width=9cm}} \caption{As  Fig.
1 for $c=100$. A zoom of the low-pressure region is presented in
the inset of each figure.}
\end{figure}

\begin{figure}[th]
\centerline{\psfig{file=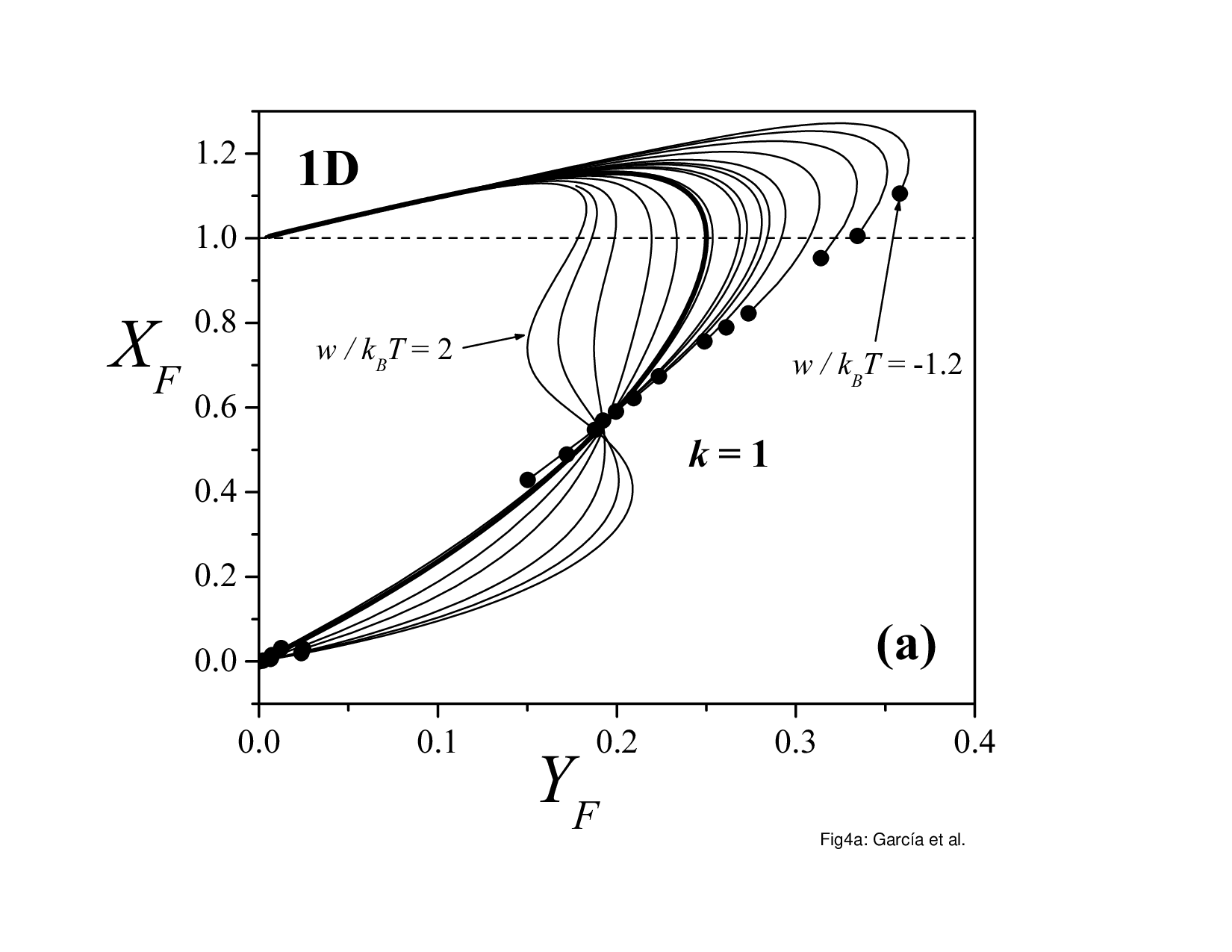,width=8.5cm}}
\centerline{\psfig{file=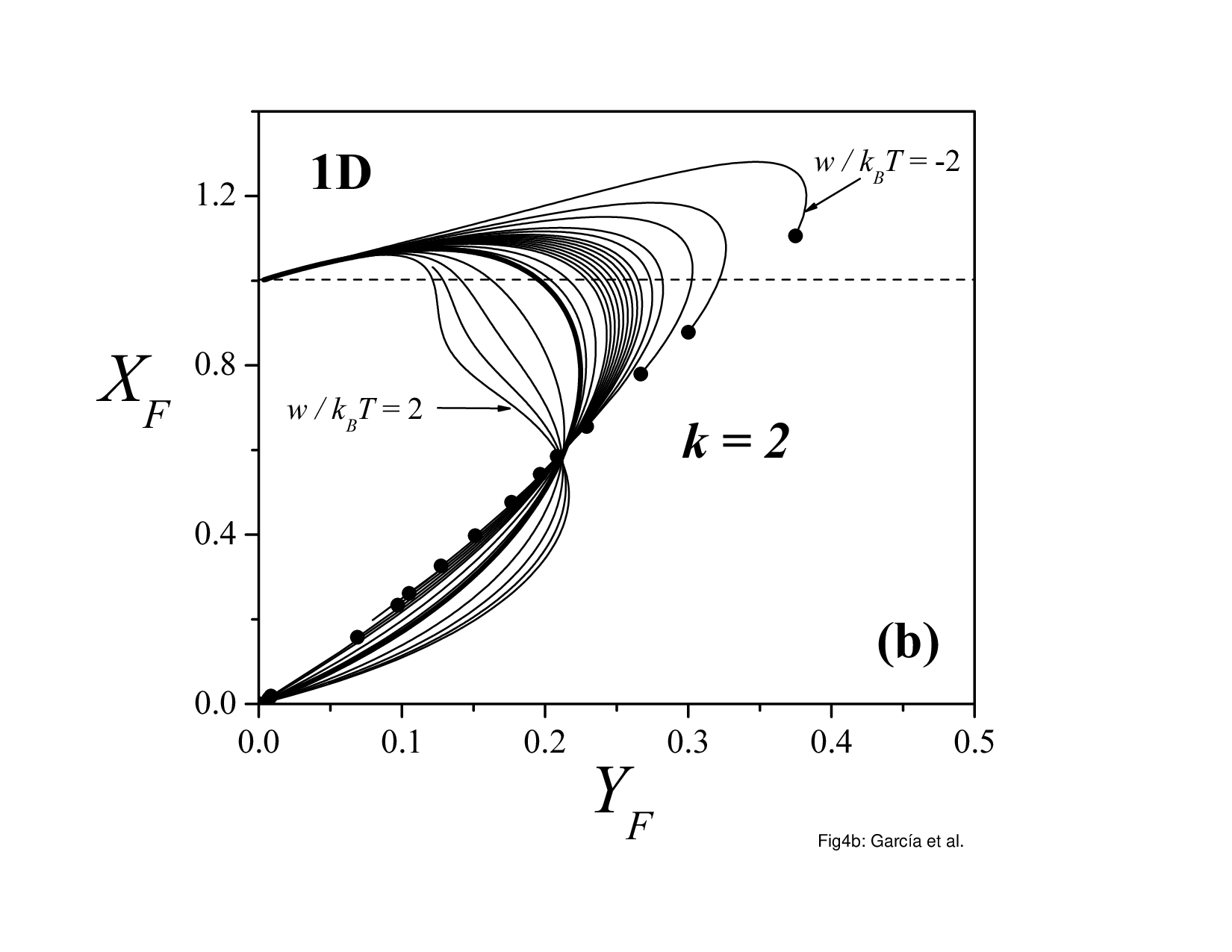,width=8.5cm}}
\centerline{\psfig{file=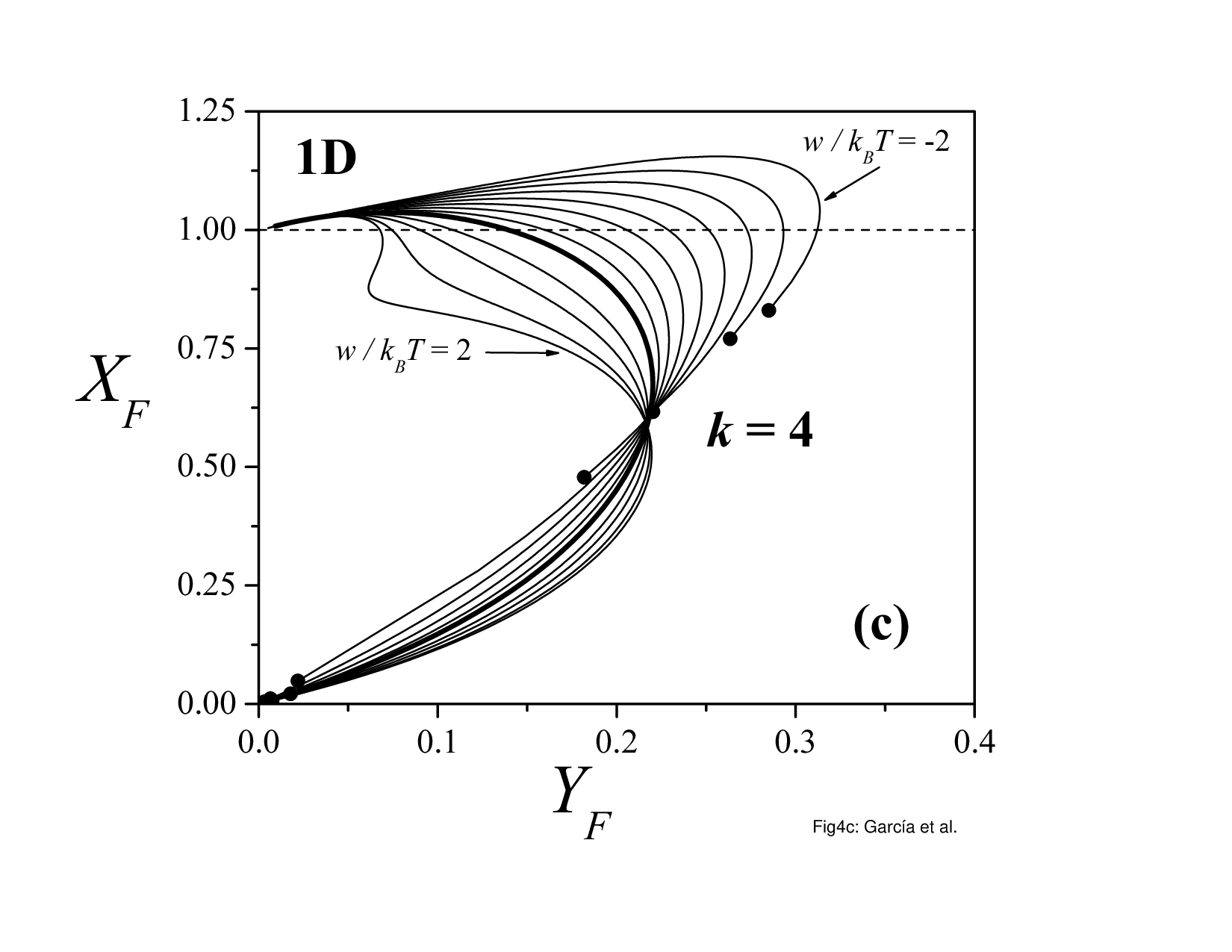,width=8.5cm}} \caption{(a)
Coordinates of the point of inflection (being $X_F$ and $Y_F$
coverage and relative pressure, respectively) for monomers ($k=1$)
adsorbed on 1D lattices and different values of $w/k_BT$
($w/k_BT=2,1.5,1.0,0.5,0.25,0.0,-0.05$,
$-0.25,-0.3,-0.4,-0.45,-0.55,-0.75,-0.95,-1.1,-1.2$). As a
reference, the curve corresponding to $w/k_BT=0$ is highlighted.
Each point on a given curve corresponds to a determined value of
$c$. Solid circles indicate the values of $c$ where the inflection
point disappears. (b) Same as in part (a) for dimers ($k=2$)
adsorbed on 1D lattices and different values of $w/k_BT$
($w/k_BT=2,1.5,1.0,0.5$,
$0.0,-0.1,-0.25,-0.4,-0.45,-0.5,-0.55,-0.6,-0.65,-0.7$,
$-0.75,-0.8,-0.9,-1.0,-1.25,-1.5,-2.0$). (c) Same as in part (a)
for tetramers ($k=4$) adsorbed on 1D lattices and different values
of $w/k_BT$ ($w/k_BT=2,1.5,1.0,0.5,0.0,-0.25,-0.5$,
$-0.75,-1.0,-1.25,-1.5,-1.75,-2.0$).}
\end{figure}

\begin{figure}[th]
\centerline{\psfig{file=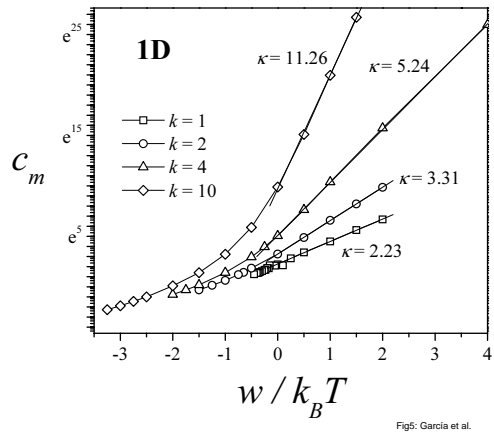,width=9cm}} \vspace*{8pt}
\caption{$\ln c_m$ (as it is indicated in the text) as a function
of $w/k_BT$ for different values of $k$ ($k=1$, $2$, $4$ and $10$)
and 1D lattices. From the slope of the curves in the range
$w/k_BT>0$ one obtains $\kappa$ (see discussion in the text).}
\end{figure}

\begin{figure}[th]
\centerline{\psfig{file=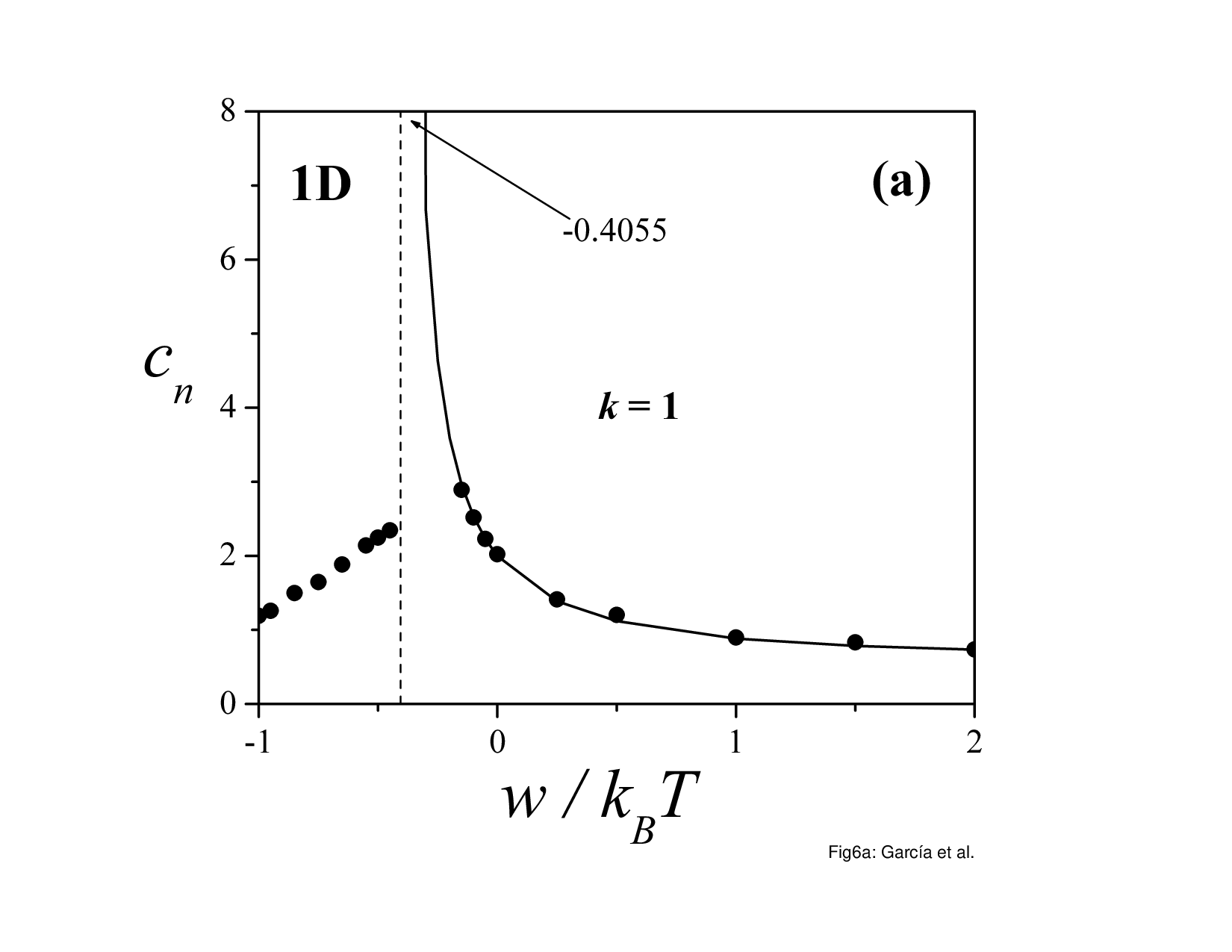,width=9cm}}
\centerline{\psfig{file=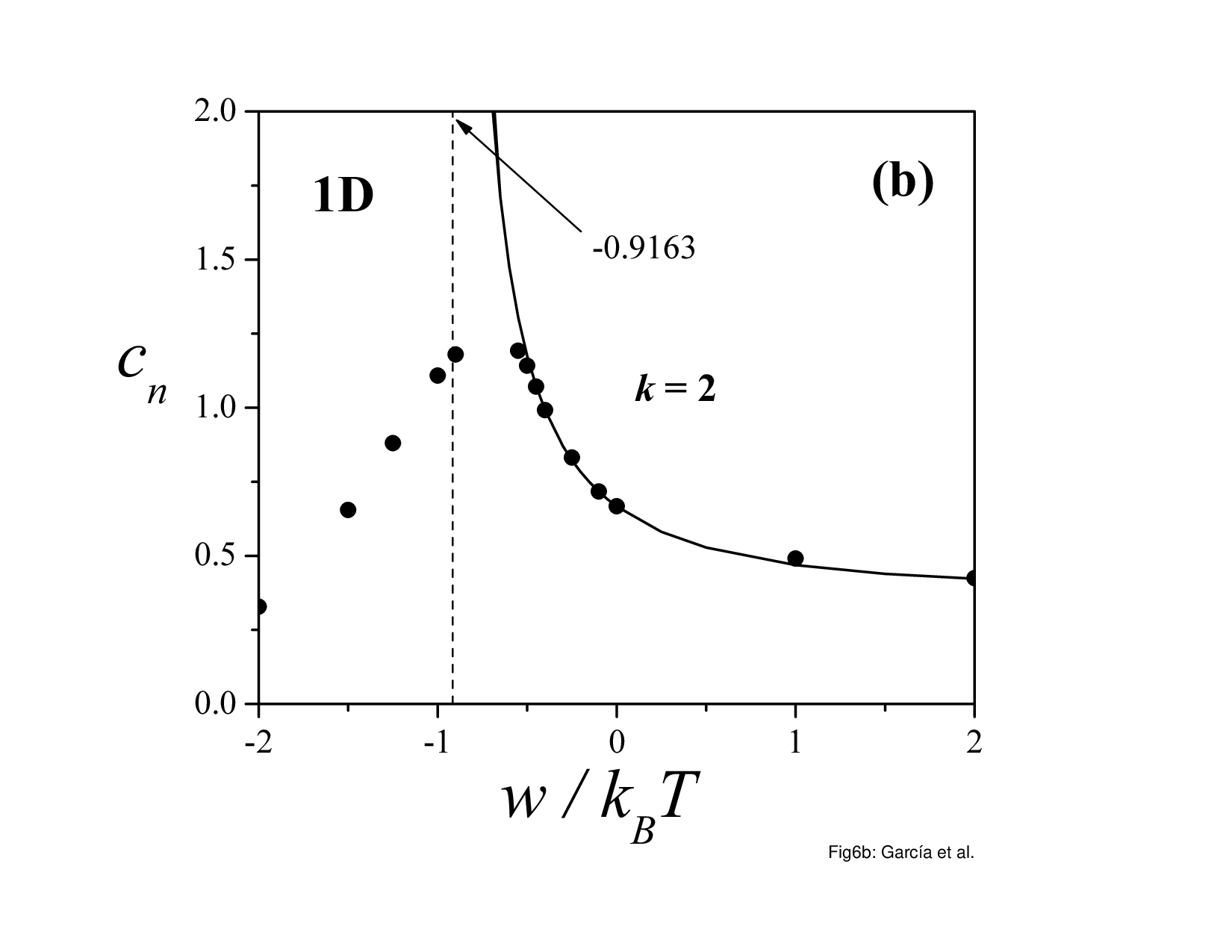,width=9cm}}
\centerline{\psfig{file=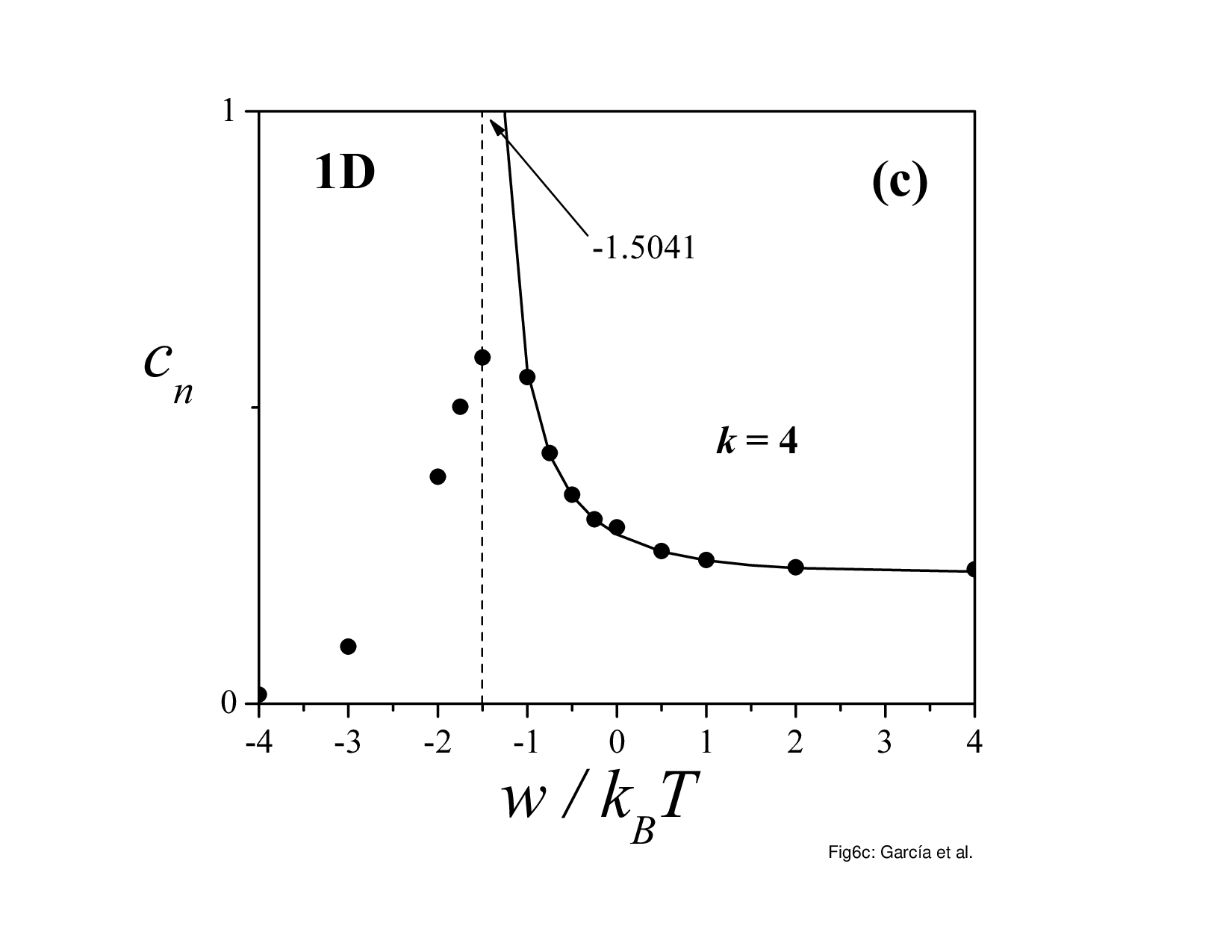,width=9cm}} \caption{(a)
$c_n$ as a function of $w/k_BT$ for $k=1$. The meaning of the
solid lines and the symbols is explained in the text. (b) As part
(a) for $k=2$ and (c) As part (a) for $k=4$.}
\end{figure}

\begin{figure}[th]
\centerline{\psfig{file=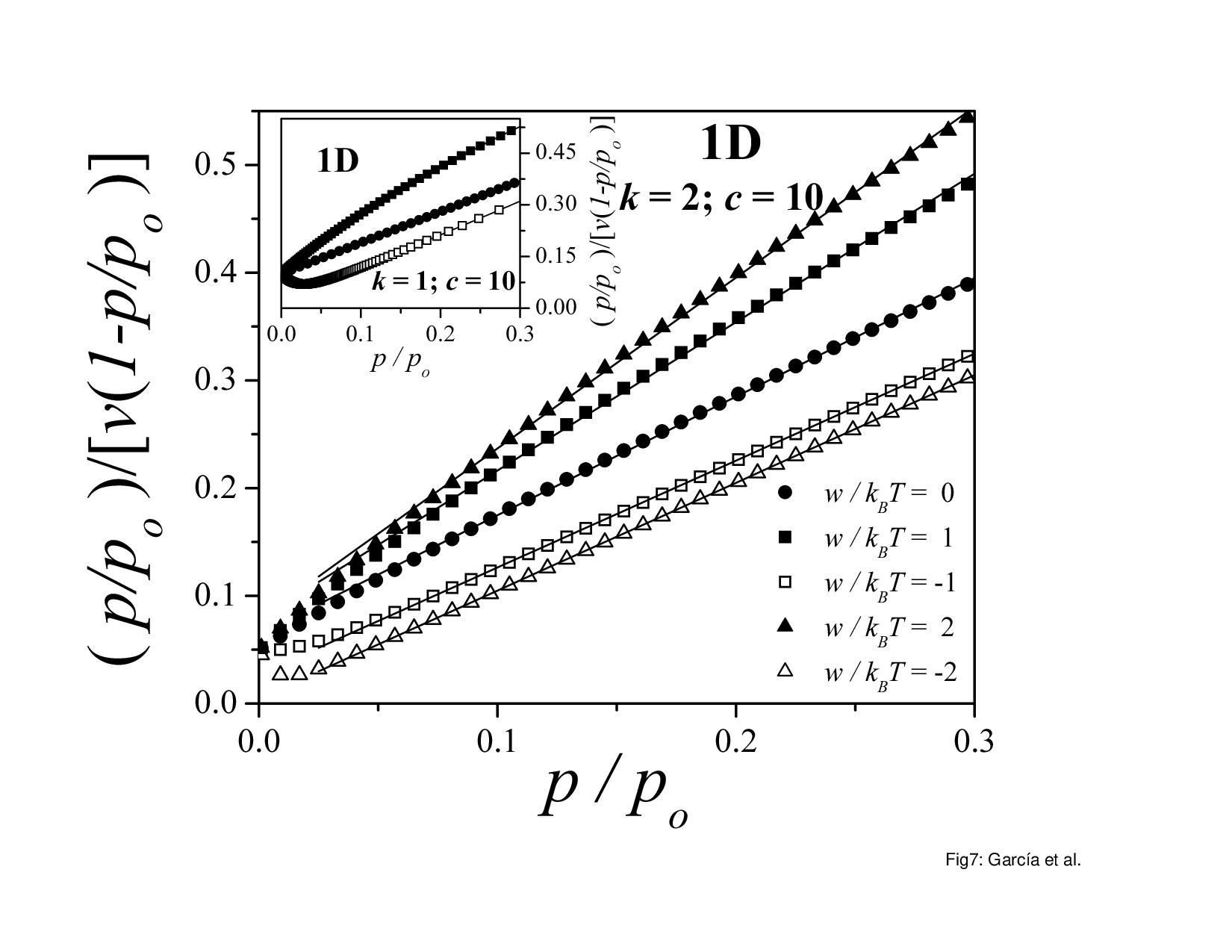,width=9cm}} \caption{$\left[\
p/p_{o} \right ] / \left [v \ (1-p/p_{o}) \right]$
 versus $p/p_{o}$ for a typical case ($c=10$ and $k=2$) and different values of $w/k_BT$
(as indicated). All curves are plotted
 in the range ($0-0.3$) of relative pressure and $v_{m}$ is set equal $1$ (in
arbitrary units). Inset: $\left[\ p/p_{o} \right ] / \left [v \
(1-p/p_{o}) \right]$ versus $p/p_{o}$ for $k=1$, $c=10$ and three
different values of $w/k_BT$: full circles, $w/k_BT=0$; open
squares, $w/k_BT=-1$ and full squares, $w/k_BT=1$.}
\end{figure}

\begin{figure}[th]
\centerline{\psfig{file=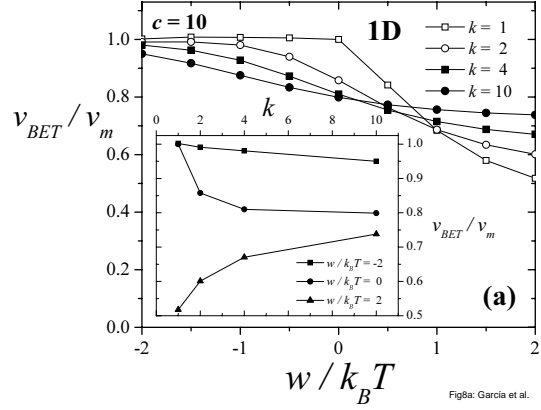,width=9cm}}
\centerline{\psfig{file=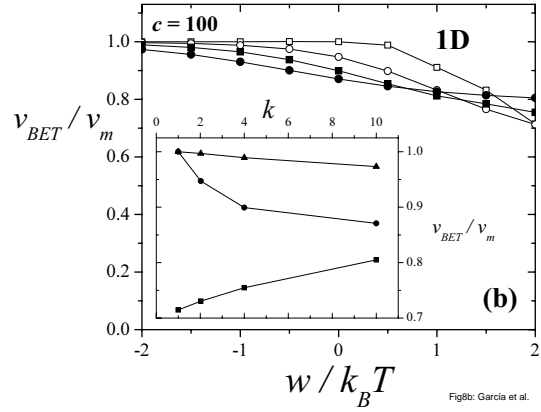,width=9cm}} \caption{(a)
Dependence on $w/k_BT$ of the monolayer volume obtained by using
BET analysis for multilayer $k$-mer adsorption on 1D lattices and
$c=10$. The curves correspond to different values of $k$ as
indicated. In the inset, the data are plotted as a function of $k$
for three values of $w/k_BT$ ($w/k_BT=-2$, $w/k_BT=0$ and
$w/k_BT=2$). (b) Same as part (a) for $c=100$.}
\end{figure}

\begin{figure}[th]
\centerline{\psfig{file=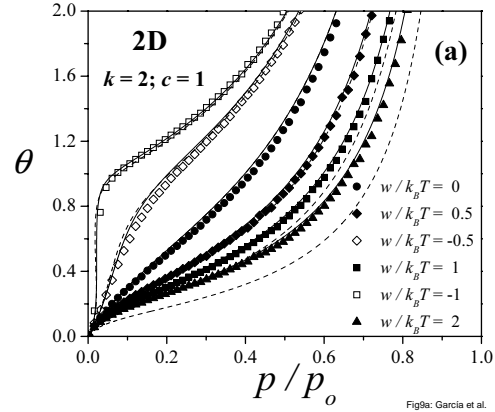,width=9cm}}
\centerline{\psfig{file=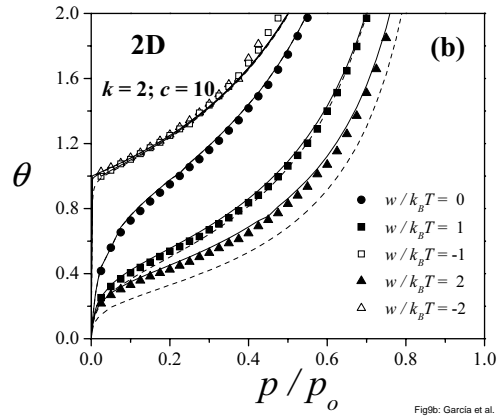,width=9cm}}
\centerline{\psfig{file=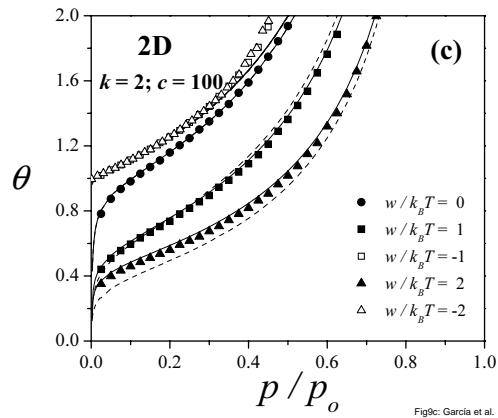,width=9cm}}
\caption{Comparison between theoretical and simulated adsorption
isotherms for dimers adsorbed on square surfaces and different
values of $w/k_BT$ (as indicated). (a) $c=1$; (b) $c=10$ and (c)
$c=100$. Symbols, solid lines and dashed lines represent results
from Monte Carlo simulations, QCA and BWA, respectively.}
\end{figure}

\begin{figure}[th]
\centerline{\psfig{file=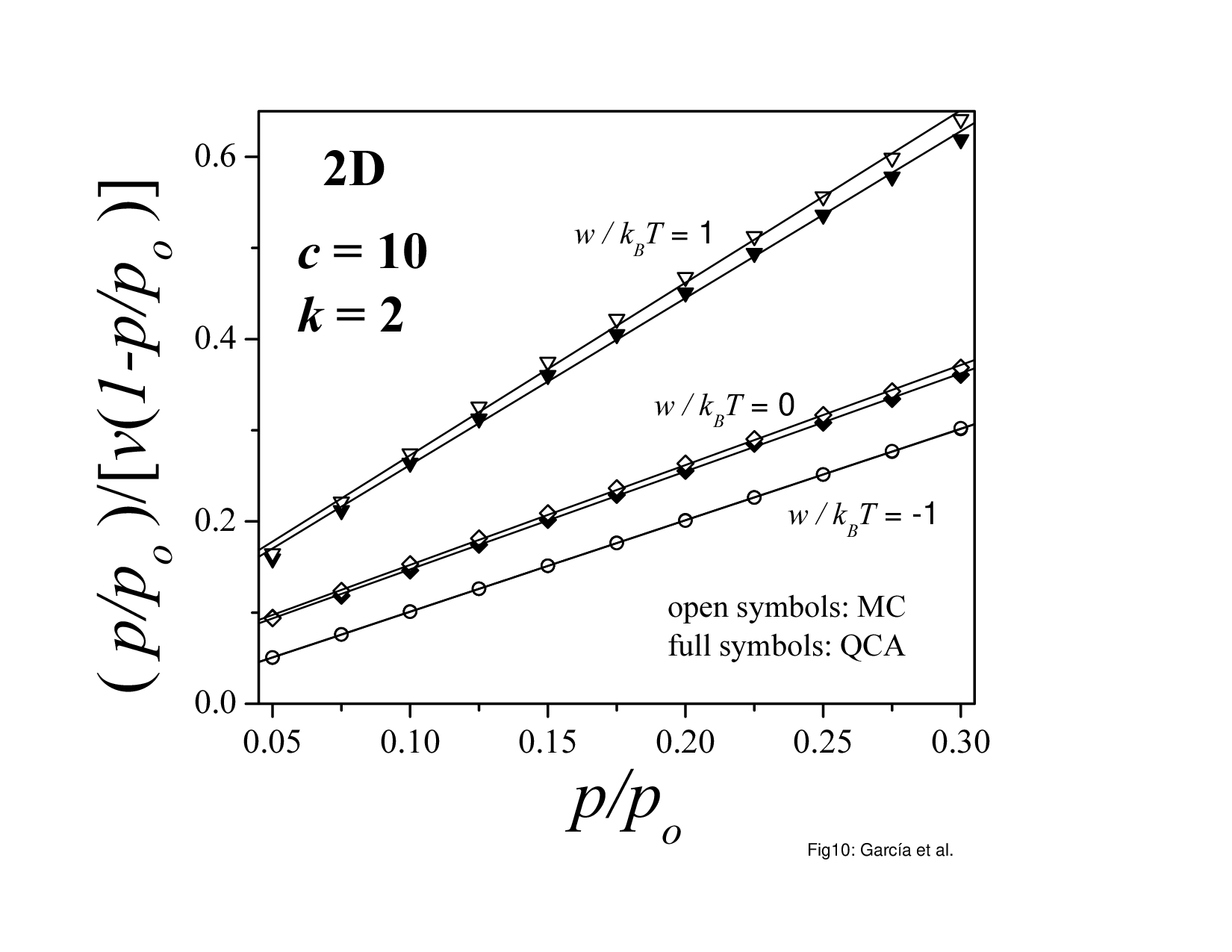,width=9cm}} \vspace*{8pt}
\caption{Comparison between theoretical and simulated adsorption
isotherms for dimers adsorbed on square surfaces with $k=2$,
$c=10$ and different values of $w/k_B T$ as indicated. The
isotherms are plotted in the range of low-relative pressure.}
\end{figure}

\begin{figure}[th]
\centerline{\psfig{file=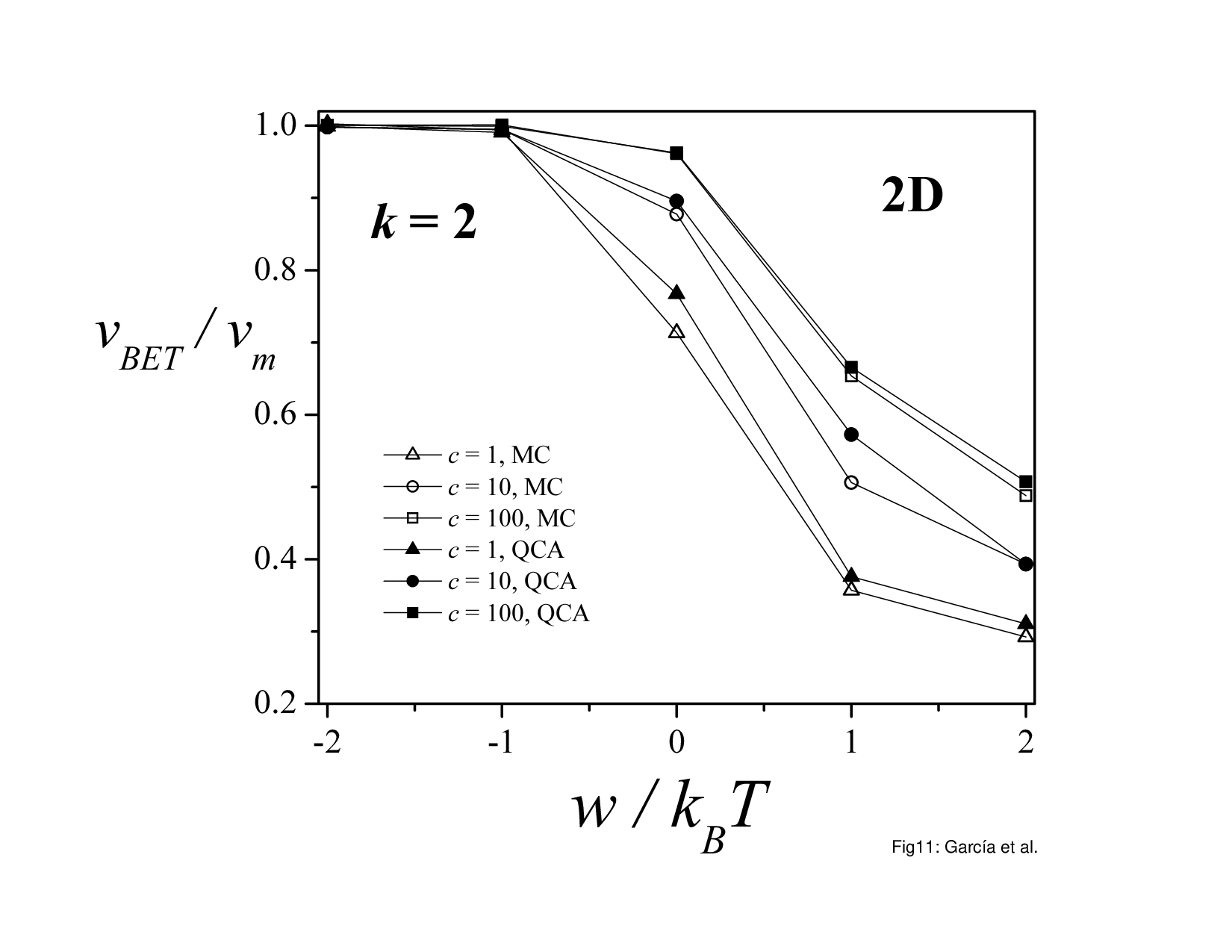,width=9cm}} \vspace*{8pt}
\caption{Dependence on $w/k_BT$ of the monolayer volume obtained
by using BET analysis for multilayer $k$-mer adsorption on square
lattices and $k=2$. Open symbols represent Monte Carlo data and
full symbols correspond to theoretical results obtained from QCA.
The curves correspond to different values of $c$ as indicated.}
\end{figure}

\begin{figure}[th]
\centerline{\psfig{file=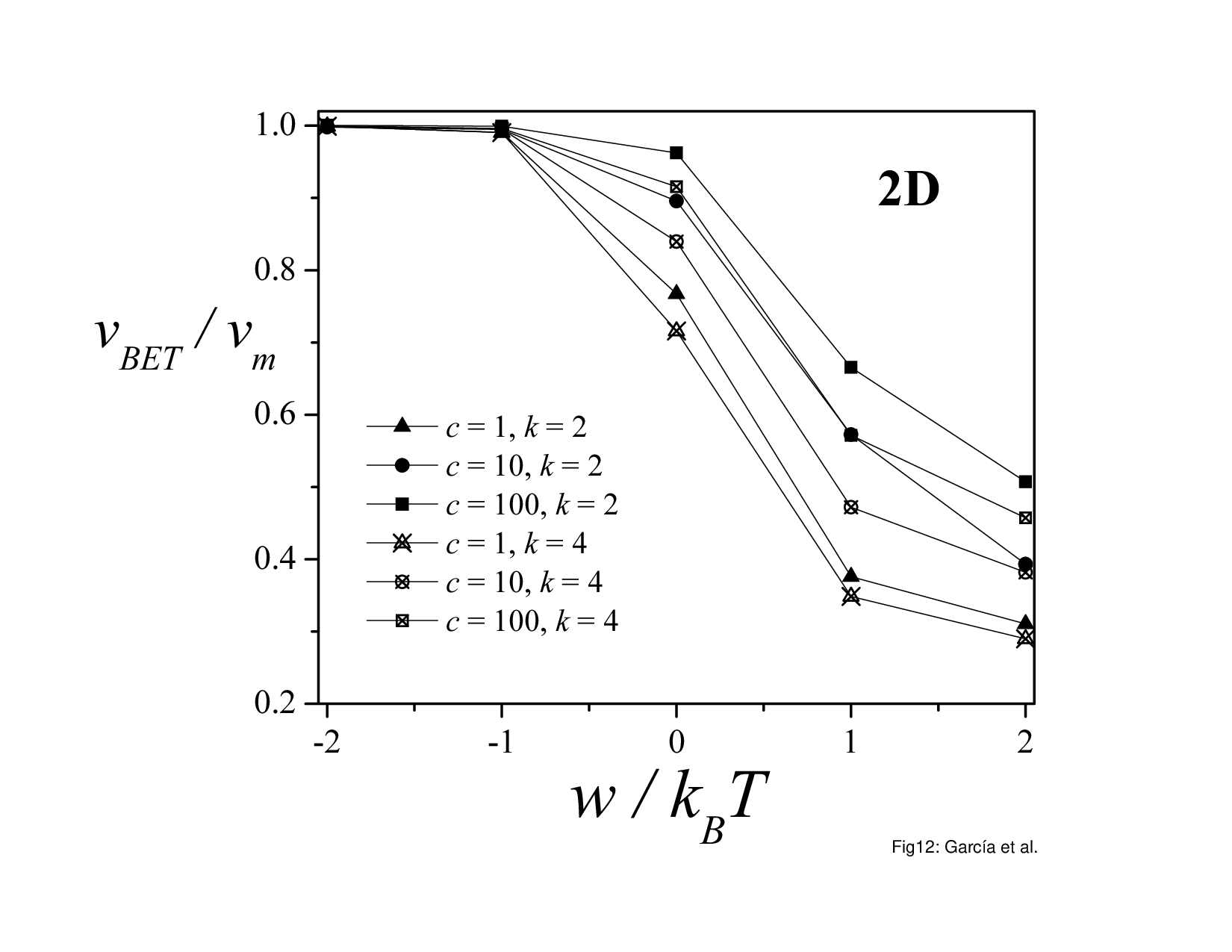,width=9cm}} \vspace*{8pt}
\caption{Dependence on $w/k_BT$ of the monolayer volume obtained
by using BET analysis for multilayer $k$-mer adsorption on square
lattices and two different values of $k$: $k=2$ (full symbols) and
$k=4$ (crossed symbols). The curves correspond to different values
of $c$ as indicated.}
\end{figure}


\begin{table}

\caption{Values of $w_{min}/k_BT$ and $\kappa$ (see discussion of
Figs. 4 and 5) for different $k$-mer sizes.}
\vspace{0.5cm}

\begin{tabular}{|c|c|c|}            \hline  \hline
$k$-mer size & $w_{min}/k_BT$ & $\kappa$ \\
\hline
$k$ = 1 & $\approx 1.1$ & $\approx 2.23$ \\
\hline
$k$ = 2 & $\approx 1.75$ & $\approx 3.31$ \\
\hline
$k$ = 4 & $\approx 2.5$ & $\approx 5.24$ \\
\hline
$k$ = 10 & $\approx 3.5$ & $\approx 11.26$ \\
\hline  \hline
\end{tabular}
\end{table}


\newpage

\begin{thebibliography}{10}

\bibitem{Hill} T. L. Hill, An Introduction to Statistical Thermodynamics,
Addison Wesley Publishing Company, Reading, MA, 1960.
\bibitem{Clark} A. Clark, The Theory of Adsorption and Catalysis, Academic Press, New York and London, 1970.
\bibitem{Steele} W. A. Steele, The interaction of gases with
solid surfaces, Pergamon Press, New York, 1974.
\bibitem{Adamson} A.W. Adamson, Physical Chemistry of Surfaces, John Wiley
  and Sons, New York, 1990.
\bibitem{Rudzi}  W. Rudzi\'nski and D. Everett, Adsorption of Gases on Heterogeneous
Surfaces, Academic Press, New York, 1992.
\bibitem{Langmuir} I. Langmuir, J. Am. Chem. Soc.  40 (1918) 1361.
\bibitem{BET} S. Brunauer, P.H. Emmet, E. Teller, J. Am. Chem. Soc.  60 (1938) 309.
\bibitem{Gregg} S.J. Gregg, K.S.W. Sing,  Adsorption, Surface Area, and Porosity, Academic Press, New York, 1991.
\bibitem{FLORY} P. J. Flory, J. Chem. Phys. 10 (1942) 51. P. J. Flory, Principles
of Polymers Chemistry, Cornell University Press, Ithaca, NY, 1953.
\bibitem{HUGGI} M. L. Huggins,  J. Phys. Chem. 46 (1942) 151. M. L. Huggins, Ann.
N.Y. Acad. Sci. 41 (1942) 1. M. L. Huggins, J. Am. Chem. Soc. 64
(1942) 1712.
\bibitem{GUGGE} E. A. Guggenheim,  Proc. R. Soc. London A183 (1944) 203.
\bibitem{DIMA} E. A. DiMarzio,  J. Chem. Phys. 35 (1961) 658.
\bibitem{Nitta} T. Nitta, M. Kuro-oka, T. Katayama, J. Chem. Eng. Jpn. 17 (1984) 45.
\bibitem{Rudzi1} W. Rudzi\'nski, K. Nieszporek, J. M. Cases, L. I.
Michot, F. Villieras, Langmuir 12 (1996) 170.
\bibitem{ARA1} G. L. Aranovich, M. D. Donohue, J. Colloid Interface Sci.
175 (1995) 492.
\bibitem{ARA2} G. L. Aranovich, M. D. Donohue, J. Colloid Interface
Sci. 189 (1997) 101.
\bibitem{LANG8} J. L. Riccardo, A. J. Ramirez-Pastor, F. Rom\'a,
Langmuir 18 (2002) 2130.
\bibitem{SS10} F. Rom\'a, A. J. Ramirez-Pastor, J. L. Riccardo,
Surf. Sci. 583 (2005) 213.
\bibitem{Ising} E. Ising, Z. Physik 31 (1925) 253.
\bibitem{Domb} C. Domb, in: C. Domb, M.S. Green (Eds.), Phase transitions and critical phenomena,
Academic Press, London-New York, 1974, Vol. 3, pp. 1-95; M.E.
Fisher, Rep. Prog. Phys. 30 (1967) 731.
\bibitem{Onsager} L. Onsager, Physical Review 65 (1944) 117.
\bibitem{Bethe} H. Bethe, Proc. R. Soc. London A150 (1935) 552.
\bibitem{Flory1} P.J. Flory, J. Chem. Phys. 10 (1942) 51.
\bibitem{Flory2} P.J. Flory, Principles of Polymers
Chemistry, Cornell University Press, Ithaca N.Y., (1953).
\bibitem{PRB3} A. J. Ramirez-Pastor, T. P. Eggarter, V. D. Pereyra,
 J. L. Riccardo, Phys. Rev. B 59 (1999) 11027.
\bibitem{LANG5} A. J. Ramirez-Pastor, J. L. Riccardo, V. Pereyra,
Langmuir 16 (2000) 10167.
\bibitem{SS12} M. D\'avila, F. Rom\'a, J. L. Riccardo, A. J. Ramirez-Pastor,
Surf. Sci. 600 (2006) 2011.
\bibitem{Metropolis}  N. Metropolis, A.W. Rosenbluth, M.N. Rosenbluth, A.H. Teller, E. Teller, J. Chem. Phys.
21 (1953) 1087.
\bibitem{LANG9} F. Rom\'a, A. J. Ramirez-Pastor, J. L. Riccardo, Langmuir 19 (2003) 6770.
\end{thebibliography}
\end{document}